\documentclass{svjour3}                     
\smartqed  
\usepackage{amssymb}
\usepackage{graphicx}
\usepackage{natbib}
\bibpunct{(}{)}{;}{a}{}{,}

\usepackage{mathptmx}      
\journalname{SSRv}
\usepackage{epsfig}

\newcommand{\be}{\begin{equation}}
\newcommand{\ee}{\end{equation}}
\newcommand{\beq}{\begin{eqnarray}}
\newcommand{\eeq}{\end{eqnarray}}

\newcommand{\hm}{\,h^{-1}{\rm Mpc}}

\newcommand{\msun}{\,h^{-1}{\rm M}_\odot}

\newcommand{\vel}{\,{\rm km\,s^{-1}}}
\newcommand{\siml}{\raise
  -2.truept\hbox{\rlap{\hbox{$\sim$}}\raise5.truept \hbox{$<$}\ }}
\newcommand{\simg}{\raise
  -2.truept\hbox{\rlap{\hbox{$\sim$}}\raise5.truept \hbox{$>$}\ }}

\begin{document}

\title{Thermodynamical properties of the ICM 
from hydrodynamical simulations}

\author{S.~Borgani \and A.~Diaferio \and K.~Dolag \and S.~Schindler}

\authorrunning{Borgani et al.}
\titlerunning{Thermodynamical properties of the ICM from simulations}

\institute{
S. Borgani \at
Department of Astronomy, University of Trieste, via Tiepolo 11, I-34143 Trieste, Italy\\
\email{borgani@oats.inaf.it}\\
INAF -- National Institute for Astrophysics, Trieste, Italy\\  
INFN -- National Institute for Nuclear Physics,
Sezione di Trieste, Italy
\and 
A. Diaferio \at
                  Dipartimento di Fisica Generale ``Amedeo Avogadro'',
                  Universit\`a degli Studi di Torino, Torino, Italy\\
                  \email{diaferio@ph.unito.it}\\ 
                   INFN -- National Institute for Nuclear Physics,
                  Sezione di Torino, Italy
\and
K. Dolag \at
                   Max-Planck-Institut f\"ur Astrophysik,
                  Karl-Schwarzschild Strasse 1, Garching bei
                  M\"unchen, Germany\\
\email{kdolag@mpa-garching.mpg.de}
\and
S. Schindler \at
Institut f\"ur Astro- und Teilchenphysik, Universit\"at
Innsbruck, Technikerstr. 25, 6020 Innsbruck, Austria\\
\email{Sabine.Schindler@uibk.ac.at}
}

\date{Received: 1 October 2007 ; Accepted: 8 November 2007 }

\maketitle

\begin{abstract}
  Modern hydrodynamical simulations offer nowadays a powerful means to
  trace the evolution of the X--ray properties of the intra--cluster
  medium (ICM) during the cosmological history of the hierarchical
  build up of galaxy clusters. In this paper we review the current
  status of these simulations and how their predictions fare in
  reproducing the most recent X--ray observations of clusters. After
  briefly discussing the shortcomings of the self--similar model,
  based on assuming that gravity only drives the evolution of the ICM,
  we discuss how the processes of gas cooling and non--gravitational
  heating are expected to bring model predictions into better
  agreement with observational data. We then present results from the
  hydrodynamical simulations, performed by different groups, and how
  they compare with observational data. As terms of comparison, we use
  X--ray scaling relations between mass, luminosity, temperature and
  pressure, as well as the profiles of temperature and entropy. The
  results of this comparison can be summarised as follows: {\em (a)}
  simulations, which include gas cooling, star formation and supernova
  feedback, are generally successful in reproducing the X--ray
  properties of the ICM outside the core regions; {\em (b)}
  simulations generally fail in reproducing the observed ``cool core''
  structure, in that they have serious difficulties in regulating
  overcooling, thereby producing steep negative central temperature
  profiles. This discrepancy calls for the need of introducing other
  physical processes, such as energy feedback from active galactic
  nuclei, which should compensate the radiative losses of the gas with
  high density, low entropy and short cooling time, which is observed
  to reside in the innermost regions of galaxy clusters.
\end{abstract}

\section{Introduction}
\label{Introduction} 
Clusters of galaxies form from the collapse of exceptionally high
density perturbations having typical size of $\sim 10$ Mpc in a
comoving frame. As such, they mark the transition between two distinct
regimes in the study of the formation of cosmic structures. The
evolution of structures involving larger scales is mainly driven by
the action of gravitational instability of the dark matter (DM)
density perturbations and, as such, it retains the memory of the
initial conditions. On the other hand, galaxy--sized structures, which
form from initial fluctuations on scales of $\sim 1$ Mpc, evolve under
the combined action of gravity and of complex gas--dynamical and
astrophysical processes. On such scales, gas cooling, star formation
and the subsequent release of energy and metal feedback from
supernovae (SN) and active galactic nuclei (AGN) have a deep impact on
the observational properties of the diffuse gas and of the galaxy
population.

In this sense, clusters of galaxies can be used as invaluable
cosmological tools and astrophysical laboratories (see \citealt{2002ARA&A..40..539R} and  \citealt{2005RvMP...77..207V}
for reviews). These two aspects are clearly interconnected with each
other. From the one hand, the evolution of the population of galaxy
clusters and their overall baryonic content provide in principle
powerful constraints on cosmological parameters. On the other hand,
for such constraints to be robust, one has to understand in detail the
physical properties of the intra--cluster medium (ICM) and its
interaction with the galaxy population.

The simplest model to predict the properties of the ICM and
their evolution has been proposed by 
\citet{1986MNRAS.222..323K}. This model is based on the assumption
that the evolution of the thermodynamical properties of the ICM is
determined only by gravity, with gas heated to the virial
temperature of the hosting DM halos by accretion shocks (\citealt{bykov2008a} - Chapter 7,
this volume). Since gravitational interaction does not introduce any
preferred scale, this model has been called
``self--similar''\footnote{Strictly speaking, self--similarity also
  requires that no characteristic scales are present in the underlying
  cosmological model. This means that the Universe must obey the
  Einstein--de-Sitter expansion law and that the shape of the power
  spectrum of density perturbations is a featureless power law. In any
  case, the violation of self--similarity introduced by the standard
  cosmological model is negligible with respect to that related to
  the non--gravitational effects acting on the gas.}. As we shall
discuss, this model provides precise predictions on the shape and
evolution of scaling relations between X--ray luminosity, entropy,
total and gas mass, which have been tested against numerical
hydrodynamical simulations (e.g., \citealt{1998ApJ...503..569E,1998ApJ...495...80B}). These predictions have been recognised
for several years to be at variance with a number of
observations. In particular, the observed relation between X--ray
luminosity and temperature (e.g. \citealt{1998ApJ...504...27M,1999MNRAS.305..631A,2004MNRAS.350.1511O}) is steeper and the measured level of gas
entropy higher than expected (e.g., \citealt{2003MNRAS.343..331P,2005A&A...429..791P}), especially for poor clusters and
groups. This led to the concept that more complex physical processes,
related to the heating from astrophysical sources of energy feedback,
and radiative cooling of the gas in the central cluster regions, play
a key role in determining the properties of the diffuse hot baryons.

Although semi--analytical approaches (e.g., 
\citealt{2001ApJ...546...63T,2005RvMP...77..207V}, and
references therein) offer invaluable guidelines to this study, it is
only with hydrodynamical simulations that one can capture the full
complexity of the problem, so as to study in detail the existing
interplay between cosmological evolution and the astrophysical
processes.

In the last years, ever improving code efficiency and supercomputing
capabilities have opened the possibility to perform simulations over
fairly large dynamical ranges, thus allowing to resolve scales of a
few kiloparsecs (kpc), which are relevant for the formation of single
galaxies, while capturing the global cosmological environment on
scales of tens or hundreds of Megaparsecs (Mpc), which are relevant for the
evolution of galaxy clusters. Starting from first attempts, in which
only simplified heating schemes were studied (e.g., \citealt{1995MNRAS.275..720N,2001ApJ...555..597B,2002MNRAS.336..409B}), a number of groups have studied the
effect of introducing also  cooling (e.g. 
\citealt{1993ApJ...412..455K,2000ApJ...536..623L,2001ApJ...552L..27M,2002ApJ...579...23D,2003MNRAS.342.1025T}), of more realistic sources of energy
feedback (e.g.,  \citealt{2004MNRAS.348.1078B,2007MNRAS.377..317K,2007astro.ph..3661N,2007arXiv0705.2238S}), of thermal conduction
(e.g., \citealt{2004ApJ...606L..97D}), and of non--thermal
pressure support from magnetic fields (e.g.,  \citealt{2001A&A...369...36D}) and cosmic rays (e.g.,  \citealt{2007MNRAS.378..385P}).

In this paper, we will review the recent advancement performed in
this field of computational cosmology and critically discuss the
comparison between simulation predictions and observations, by
restricting the discussion to the thermal effects. As such, this
paper complements the reviews by \citealt{borgani2008} - Chapter 18,
this volume, which reviews the study of the ICM chemical enrichment
and by \citealt{dolag2008b} - Chapter 15, this volume, which reviews
the study of the non--thermal properties of the ICM from
simulations. We refer to the reviews by \citealt{dolag2008a} - Chapter
12, this volume for a description of the techniques of numerical
simulations and by \citealt{kaastra2008} - Chapter 9, this volume for
an overview of the observed thermal properties of the ICM.

The scheme of the presentation is as follows. In Sect. \ref{models}
we  briefly discuss the self--similar model of the ICM and how the
action of non--gravitational heating and cooling are expected to alter
the predictions of this model. Sect.~\ref{scalings} and
\ref{profiles}  overview the results obtained on the comparison
between observed and simulated scaling relations and profiles of
X--ray observable quantities, respectively. In Sect.~\ref{summary}
we  summarise and critically discuss the results presented.

\section{Modelling the ICM}
\label{models}
\subsection{The self--similar scaling}
The simplest model to predict the observable properties of the ICM is
based on the assumption that gravity only determines the
thermodynamical properties of the hot diffuse gas
\citep{1986MNRAS.222..323K}. Since gravity does not have preferred
scales, we expect clusters of different sizes to be the scaled version
of each other. This is the reason why this model has been called
self-similar.

If, at redshift $z$, we define $M_{\Delta_{\displaystyle{\rm c}}}$ to be the mass contained
within the radius $r_{\Delta_{\displaystyle{\rm c}}}$, encompassing a mean density
$\Delta_{\rm c}$ times the critical density $ \rho_{\rm c}(z) $, then
$M_{\Delta_{\displaystyle{\rm c}}} \propto \rho_{\rm c}(z) \Delta_{\rm c} r_{\Delta_{\displaystyle{\rm c}}}^3 $. 
The critical density of the universe scales with
redshift as $\rho_{\rm c}(z)=\rho_{\rm c0} E^2(z)$, where $E(z)$ is given by
\be
E(z)\,=\,\left[(1+z)^3\Omega_{\rm m}+(1+z)^2\Omega_{\rm k}+\Omega_\Lambda\right]^{1/2}\,,
\ee
where $\Omega_{\rm m}$ and $\Omega_\Lambda$ are the density parameters
associated to the non--relativistic matter and to the cosmological
constant, respectively, $\Omega_{\rm k}=1-\Omega_{\rm m}-\Omega_\Lambda$ and we
neglect any contribution from relativistic species.

Therefore, the cluster size $r_{\Delta_{\displaystyle{\rm c}}}$ scales with $z$ and
$M_{\Delta_{\displaystyle{\rm c}}}$ as $r_{\Delta_{\displaystyle{\rm c}}}\propto M_{\Delta_{\displaystyle{\rm c}}}^{1/3}
E^{-2/3}(z)$, so that, assuming hydrostatic equilibrium, cluster
mass scales with temperature $T$ as
\begin{equation}
M_{\Delta_{\displaystyle{\rm c}}}\,\propto \,T^{3/2}E^{-1}(z)\,.
\label{eq:mt_ss}
\end{equation}
If $\rho_{\rm gas}$ is the gas density, the corresponding 
X--ray luminosity is
\begin{equation}
L_X\,=\, \int_V \left({\rho_{\rm gas}\over \mu m_{\rm p}}\right)^2
\Lambda(T)\,{\rm d}V\,, 
\label{eq:lx}
\end{equation}
where $\Lambda(T)\propto T^{1/2}$ for pure thermal Bremsstrahlung
emission. If gas accretes along with DM by gravitational
instability during the formation of the cluster halo, then we expect
that $\rho_{\rm gas}(r)\propto \rho_{\rm DM}(r)$, so that
\begin{equation}
L_X\,\propto\,M_{\Delta_{\displaystyle{\rm c}}}\rho_{\rm c} T^{1/2}\,\propto \, T^2 E(z)\,.
\label{eq:lt_ss}
\end{equation}

Another useful quantity characterising the thermodynamical properties
of the ICM is the entropy \citep{2005RvMP...77..207V} which, in X--ray
studies of the ICM, is usually defined as
\be
K\,=\,{{\rm k}_{\rm B}T\over \mu m_{\rm p} \rho_{\rm gas}^{2/3}}\,,
\label{eq:entrK} 
\ee
where ${\rm k}_{\rm B}$ is the Boltzmann constant, $\mu$ the mean molecular weight
($\simeq 0.58$ for a plasma of primordial composition) and $m_{\rm p}$ the
proton mass. With the above definition, the quantity $K$ is the
constant of proportionality in the equation of state of an adiabatic
mono-atomic gas, $P=K\rho_{\rm gas}^{5/3}$. Using the thermodynamic
definition of specific entropy, $s=c_V\ln(P/\rho_{\rm gas}^{5/3})$ ($c_V$:
heat capacity at constant volume), one obtains $s={\rm k}_{\rm B}\ln
K^{3/2}+s_0$, where $s_0$ is a constant. Another quantity, often
called ``entropy'' in the cluster literature, which we will also use
in the following, is
\be
S\,=\,{\rm k}_{\rm B} T n_{\rm e}^{-2/3}\,,
\label{eq:entr}
\ee 
where $n_{\rm e}$ is the electron number density. According to the
self--similar model, this quantity, computed at a fixed overdensity
$\Delta_{\rm c}$, scales with temperature and redshift according to
\be
S_{\Delta_{\displaystyle{\rm c}}}\propto T (1+z)^{-2}\,.
\label{eq:entr_ss}
\ee

As already mentioned in the introduction, a number of observational
facts from X--ray data point against the simple self--similar
picture. The steeper slope of the $L_X$--$T$ relation
\citep{1998ApJ...504...27M,1999MNRAS.305..631A,2004MNRAS.350.1511O},
$L_X\propto T^\alpha$ with $\alpha \simeq 3$ for clusters and possibly
larger for groups, the excess entropy in poor clusters and groups
\citep{2003MNRAS.343..331P,2005A&A...429..791P,2005A&A...433..101P} and
the decreasing trend of the gas mass fraction in poorer systems
\citep{2003ApJ...591..749L,2003MNRAS.340..989S} all point toward the
presence of some mechanism which significantly affects the ICM
thermodynamics.

\subsection{Heating and cooling the ICM}
\label{s:heatcool}
The first mechanism, that has been introduced to break the ICM
self--similarity, is non--gravitational heating (e.g. 
\citealt{1991ApJ...383...95E,1991ApJ...383..104K,2001ApJ...546...63T}). The idea is that by
increasing the gas entropy with a given extra heating energy per gas
particle $E_{\rm h}$ prevents gas from sinking to the centre of DM
halos, thereby reducing gas density and X--ray emissivity. This effect
will be large for small systems, whose virial temperature is
${\rm k}_{\rm B}T\lesssim E_{\rm h}$, while leaving rich clusters with ${\rm k}_{\rm B}T\gg E_{\rm h}$
almost unaffected. Therefore, we expect that the X--ray luminosity and
gas content are relatively more suppressed in poorer systems, thus
leading to a steepening of the $L_X$--$T$ relation.

\begin{figure}
\centerline{
\psfig{figure=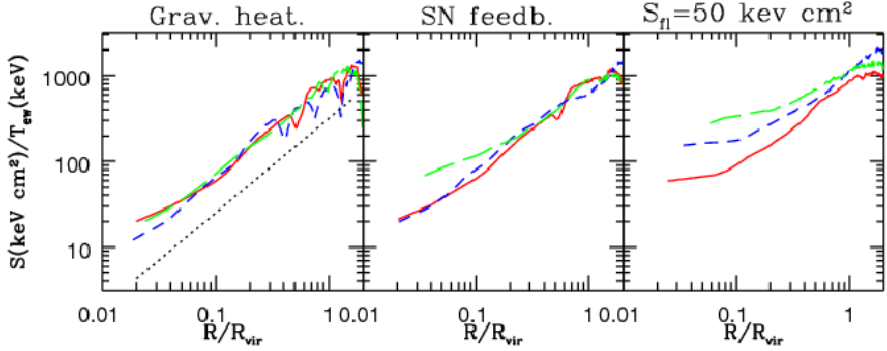,height=5cm}
}
\caption{Profiles of reduced entropy, $S/T$, for
  non--radiative simulations (from  \protect\citealt{2001ApJ...559L..71B}). The left panel is for
  simulations including only gravitational heating, the central panel
  is for runs including SN feedback as predicted by a semi--analytical
  model of galaxy formation and the right panel is with pre--heating
  with an entropy threshold at redshift $z=3$. Solid, short--dashed
  and long--dashed curves are for a 3 keV cluster, for a 1 keV group
  and for a 0.5 keV group, respectively.  The dotted straight line in
  the left panel shows the analytical prediction by \protect\citet{2001ApJ...546...63T} for the entropy profile
  associated to gravitational heating.}
\label{fi:borg01}
\end{figure}

The notion of non--gravitational heating has been first implemented in
non--radiative (i.e. neglecting the effect of cooling) hydrodynamical
simulations by either injecting entropy in an impulsive way at a given
redshift \citep{1995MNRAS.275..720N,2001ApJ...555..597B}, or by adding
energy in a redshift--modulated way, so as to mimic the rate of SN
explosions from an external model of galaxy formation
\citep{2002MNRAS.336..409B}. In Fig.~\ref{fi:borg01} we show the
different efficiency that different heating mechanisms have in
breaking the self--similar behaviour of the entropy profiles in
objects of different mass, ranging from a Virgo--like cluster to a
poor galaxy group. According to the self--similar model, the profiles
of reduced entropy, $S/T$, should be independent of the cluster
mass. This is confirmed by the left panel, which also shows that
these profiles have a slope consistent with that predicted by a model
in which gas is shock heated by spherical accretion in a DM halo,
under the effect of gravity only \citep{2001ApJ...546...63T}. The
central panel shows instead the effect of adding energy from SN, whose
rate is that predicted by a semi--analytical model of galaxy
formation. In this case, which corresponds to a total heating energy of
about 0.3 keV/particle, the effect of extra heating starts being visible,
but only for the smaller system. It is only with the pre--heating
scheme, based on imposing an entropy floor of 50 keV cm$^2$, that
self--similarity is clearly broken. While this heating scheme is
effective in reproducing the observed $L_X$--$T$ relation, it produces
large isentropic cores, a prediction which is at variance with respect
to observations (e.g., \citealt{2006ApJ...643..730D}).

\begin{figure}
\centerline{
\hbox{
\psfig{figure=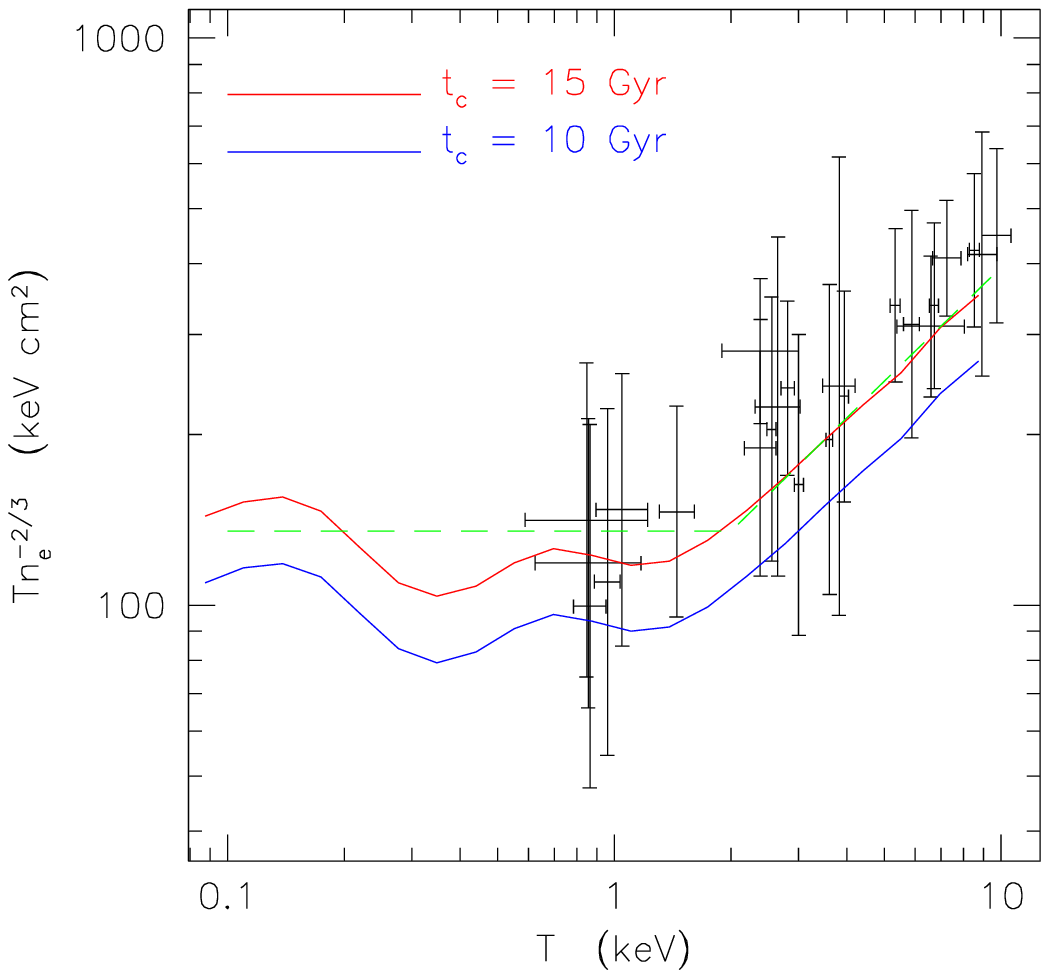,height=5.5cm}
\psfig{figure=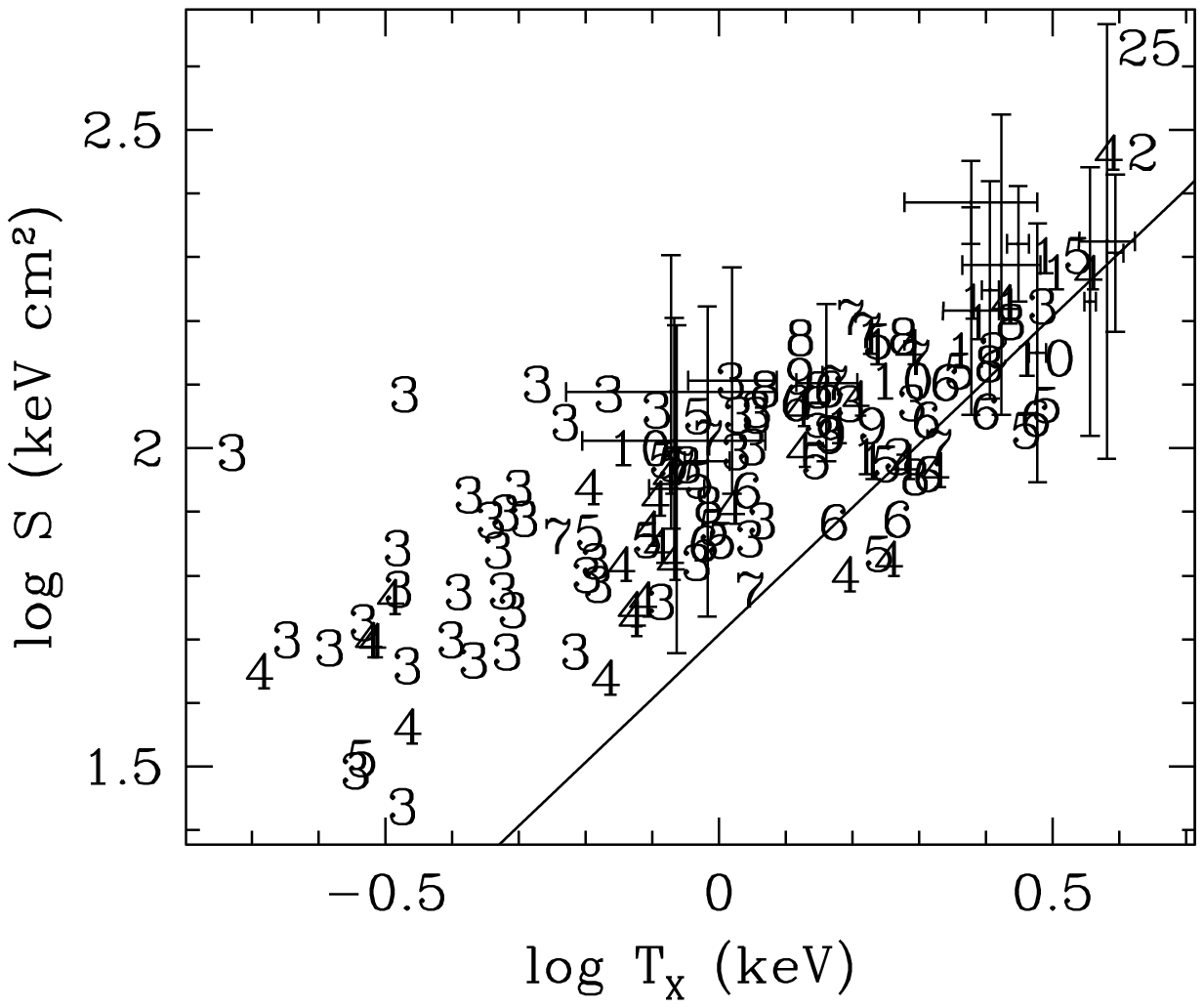,height=6.5cm}
}}
\caption{Left panel: the relation between entropy and temperature for
  gas having a fixed value of the cooling time (from 
  \protect\citealt{2001Natur.414..425V}). The crosses with error bars are
  observational data on the entropy measured at one--tenth of the
  virial radius for clusters and groups by 
  \protect\citet{1999Natur.397..135P}. Right panel: a comparison
  between observations (error bars with crosses;  \protect\citet{1999Natur.397..135P} and simulations, including
  radiative cooling and star formation (numbers), for the entropy in
  the central regions of galaxy groups and clusters. The solid line
  shows the prediction of the self--similar model (from  \protect\citealt{2002ApJ...579...23D}).}
\label{fi:dave}
\end{figure}

Although it may look like a paradox, radiative cooling has been also
suggested as a possible alternative to non--gravitational heating to
increase the entropy level of the ICM and suppressing the gas content
in poor systems. As originally suggested by \citet{2001Natur.414..425V}, cooling provides a selective removal of
low--entropy gas from the hot X--ray emitting phase (see also  \citealt{2002ApJ...572L..19W}). As a consequence, while the global
entropy of the baryons decreases, the entropy of the X--ray emitting
gas increases. This is illustrated in the left panel of Fig.~\ref{fi:dave} (from \citealt{2001Natur.414..425V}). In
this plot, each of the two curves separates the upper portion of the
entropy--temperature plane, where the gas has cooling time larger than
the age of the system, from the lower portion, where gas with short
cooling time resides. This implies that only gas having a relatively
high entropy will be observed as X--ray emitting, while the
low--entropy gas will be selectively removed by radiative cooling.
The comparison with observational data of clusters and groups, also
reported in this plot, suggests that their entropy level may well be
the result of this removal of low--entropy gas operated by radiative
cooling. This analytical prediction has been indeed confirmed by
radiative hydrodynamical simulations. The right panel of
Fig.~\ref{fi:dave} shows the results of the simulations by \citet{2002ApJ...579...23D} on the temperature dependence of the
central entropy of clusters and groups. Quite apparently, the entropy
level in simulations is well above the prediction of the self--similar
model, by a relative amount which increases with decreasing
temperature, and in reasonable agreement with the observed entropy
level of poor clusters and groups.

Although cooling may look like an attractive solution, it suffers from
the drawback that a too large fraction of gas is converted into stars
in the absence of a source of heating energy which regulates the
cooling runaway. Indeed, while observations indicate that only about
10 per cent of the baryon content of a cluster is in the stellar phase
(e.g., \citealt{2001MNRAS.326.1228B,2003ApJ...591..749L}), radiative simulations, like those
shown in Fig.~\ref{fi:dave}, convert into stars up to $\sim 50$ per
cent of the gas.

\begin{figure}
\hbox{
\psfig{figure=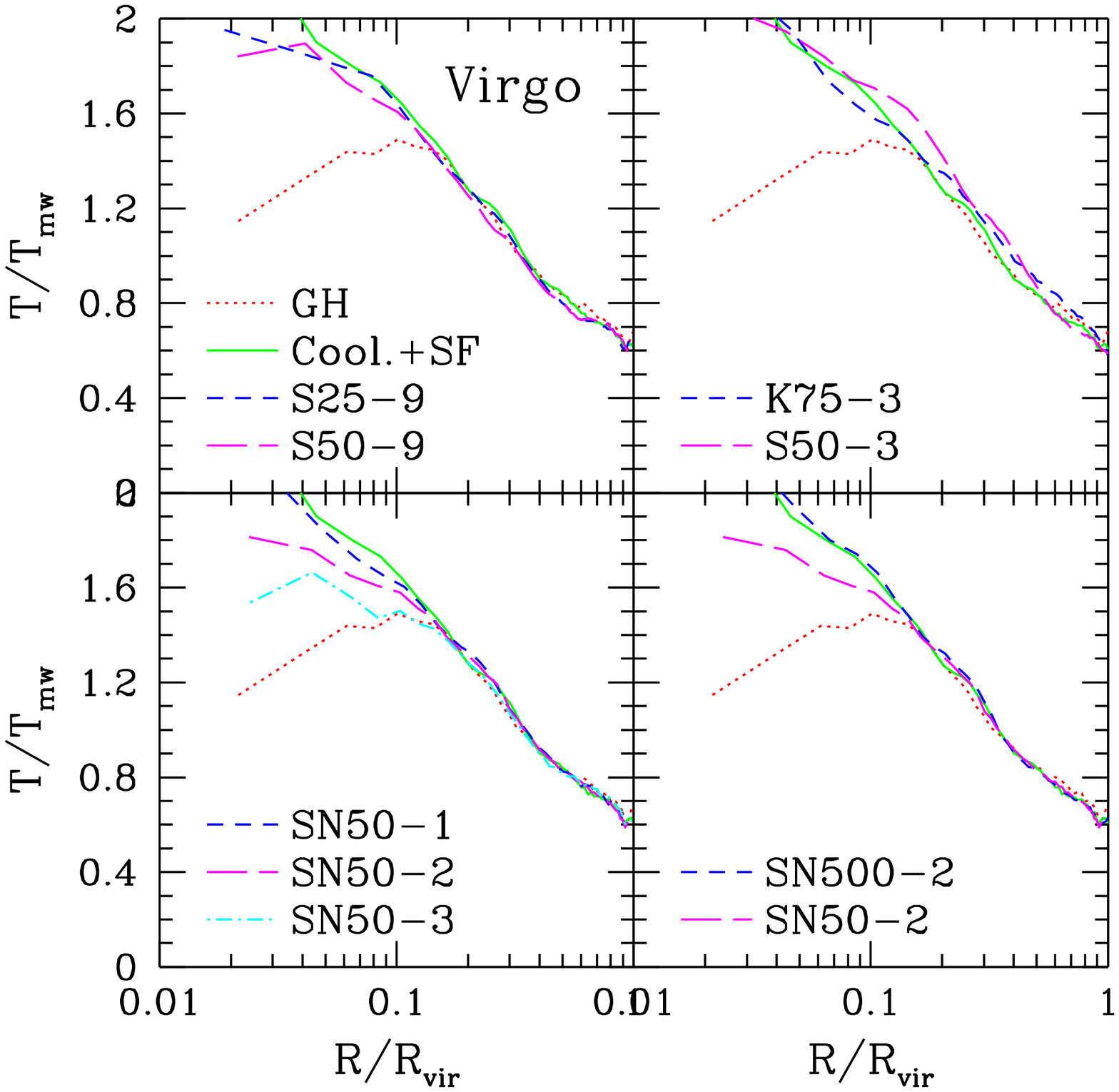,height=6.5cm}
\psfig{figure=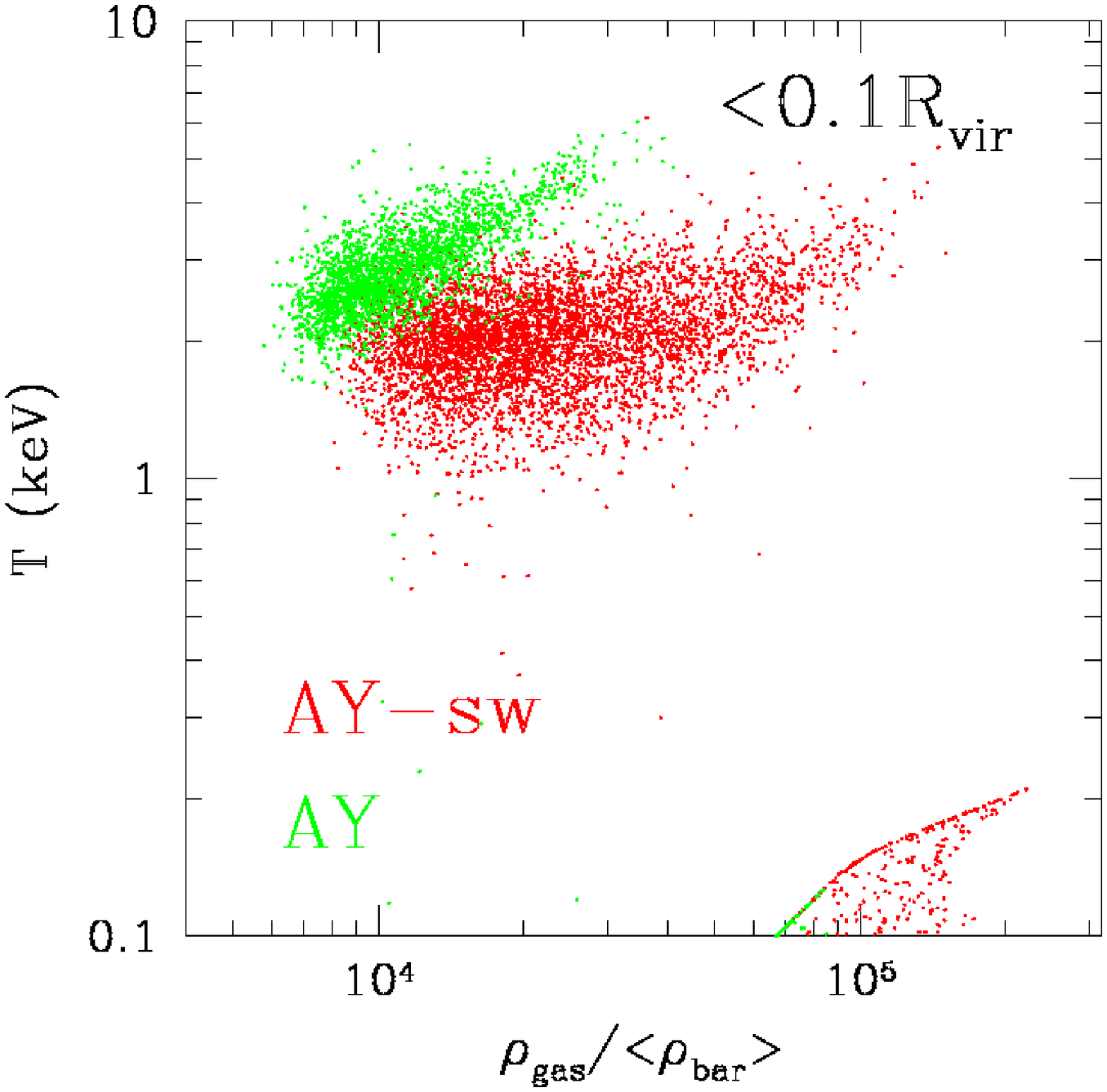,height=6.5cm}
}
\caption{Left panel: temperature profiles from hydrodynamical
  simulations of a $\sim 3$ keV galaxy cluster. In all panels the
  dotted and the solid curves correspond to a non radiative run and to
  a run including cooling and star formation. The other curves are for
  different recipes of gas heating (from 
  \protect\citealt{2003MNRAS.342.1025T}). Right panel: the relation
  between temperature and overdensity for gas particles within
  0.1$r_{200}$ for SPH simulations of a cluster of mass $\simeq
  10^{14}\msun$. Upper (green) points are for a run which includes
  feedback through galactic winds with a velocity of $500~\vel$, while
  the lower (red) points are for a run based on assuming stronger
  winds (sw), with a twice as large velocity. Both runs include a
  model of chemical enrichment (\citealt{borgani2008} - Chapter 18, this
  volume) which assumes an Initial Mass Function for star formation by
  \protect\citealt{1987A&A...173...23A} (AY). The
  points in the bottom right corner are star--forming gas particles.}
\label{fi:tsim}
\end{figure}

Another paradoxical consequence of cooling is that it increases the
temperature of the hot X--ray emitting gas at the centre of
clusters. This is shown in the left panel of Fig.~\ref{fi:tsim}
(from \citealt{2003MNRAS.342.1025T}), which compares
the temperature profiles for the non--radiative run of a Virgo--like
cluster with a variety of radiative runs, based on different ways of
supplying non--gravitational heating. The effect of introducing
cooling is clearly that of steepening the temperature profiles in the
core regions, while leaving it unchanged at larger radii. The reason
for this is that cooling causes a lack of central pressure support.
As a consequence, gas starts flowing in sub-sonically from more
external regions, thereby being heated by adiabatic compression. As we
shall discuss in Sect.~\ref{profiles}, this feature of cooling makes
it quite difficult to reproduce the structure of the cool cores
observed in galaxy clusters.

Steepening of the central temperature profiles and overcooling are two
aspects of the same problem. In principle, the solution to this
problem should be provided by a suitable scheme of gas heating which
regulates star formation, while maintaining pressurised gas in the hot
phase. The right panel of Fig.~\ref{fi:tsim} compares the
temperature--density phase diagrams for gas particles lying in the
central region of an SPH--simulated cluster, when using two different
feedback efficiencies. The two simulations include cooling, star
formation and feedback in the form of galactic winds powered by SN
explosions, following the scheme introduced by \citet{2003MNRAS.339..289S}. The upper (green) and the lower (red)
clouds of high temperature particles correspond to a wind velocity of
$500\vel$ and of $1000\vel$, respectively. This plot illustrates
another paradoxical effect: in the same way that cooling causes an
increase of the temperature of the hot phase, supplying energy with an
efficient feedback causes a decrease of the temperature. The reason
for this is that extra energy compensates radiative losses, thereby
maintaining the pressure support for gas which would otherwise have a
very short cooling time, thereby allowing it to survive on a lower
adiabat. It is also worth reminding that cooling efficiency increases
with the numerical resolution (e.g., \citealt{2001MNRAS.326.1228B,2006MNRAS.367.1641B}). Therefore, for a feedback mechanism
to work properly, it should be able to stabilise the cooling
efficiency in a way which is independent of resolution.

In the light of these results, it is clear that the observed lack of
self--similarity in the X--ray properties of clusters cannot be simply
explained on the grounds of a single effect. The emerging picture is
that the action of cooling and of feedback energy, e.g. associated to
SN explosions and AGN, should combine in a self--regulated way. As we
shall discuss in the following sections, hydrodynamical simulations of
galaxy clusters in a cosmological context demonstrate that achieving
this heating/cooling balance is not easy and represents nowadays one
of the most challenging tasks in the numerical study of clusters.

\begin{figure*}
\centerline{
\vbox{
\hbox{
\psfig{file=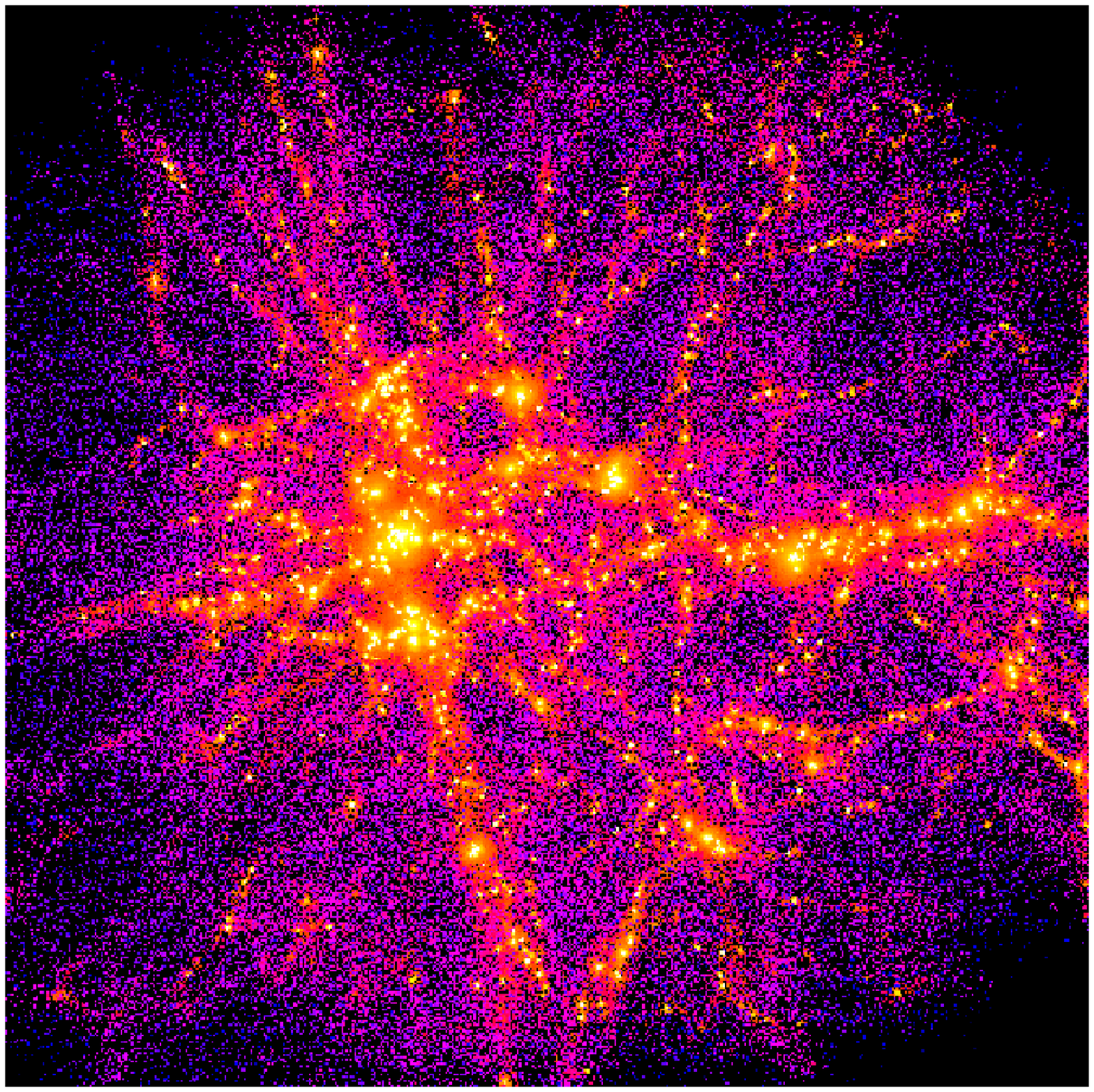,width=6.2cm} 
\psfig{file=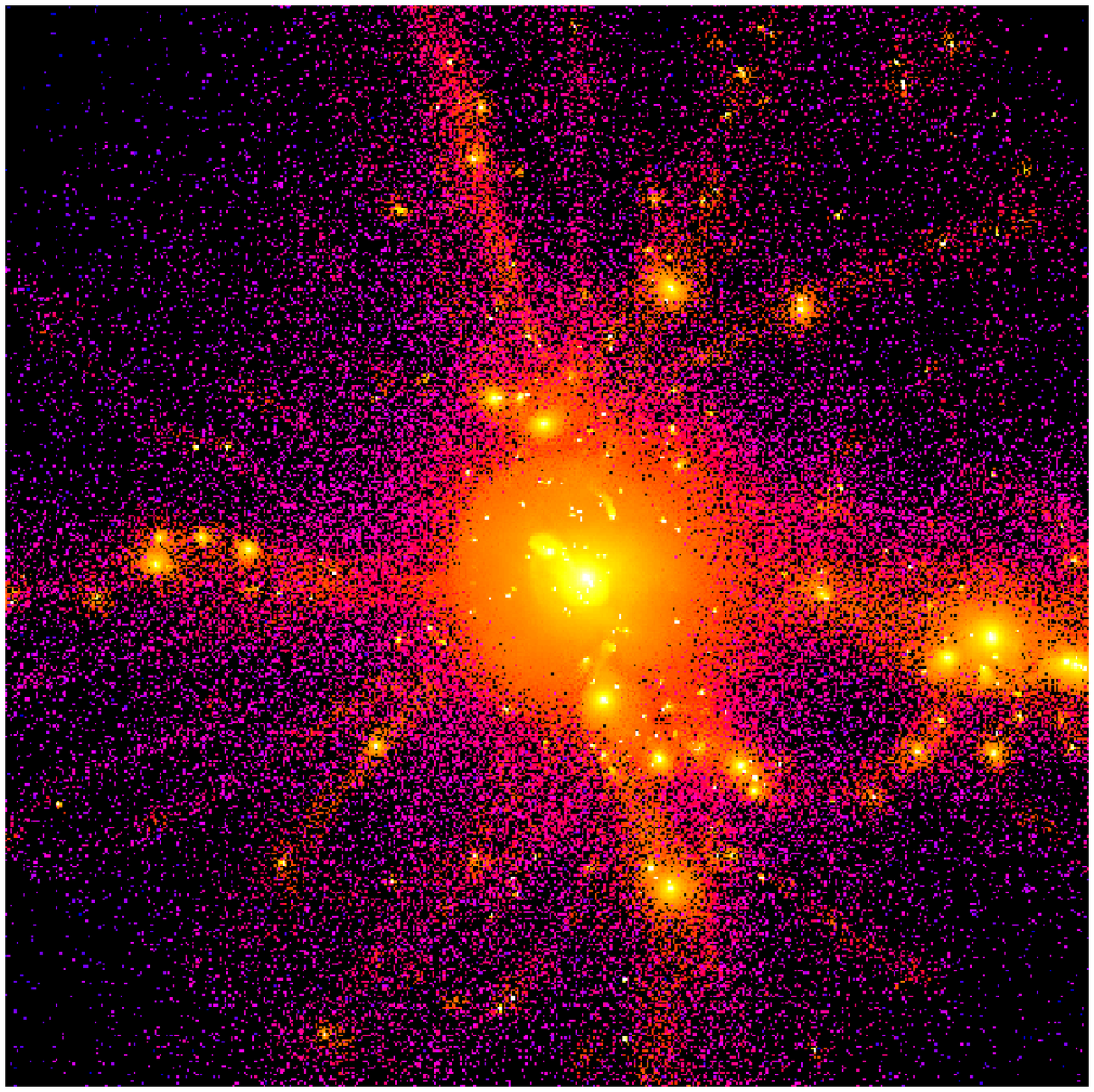,width=6.2cm} 
}
\hbox{
\psfig{file=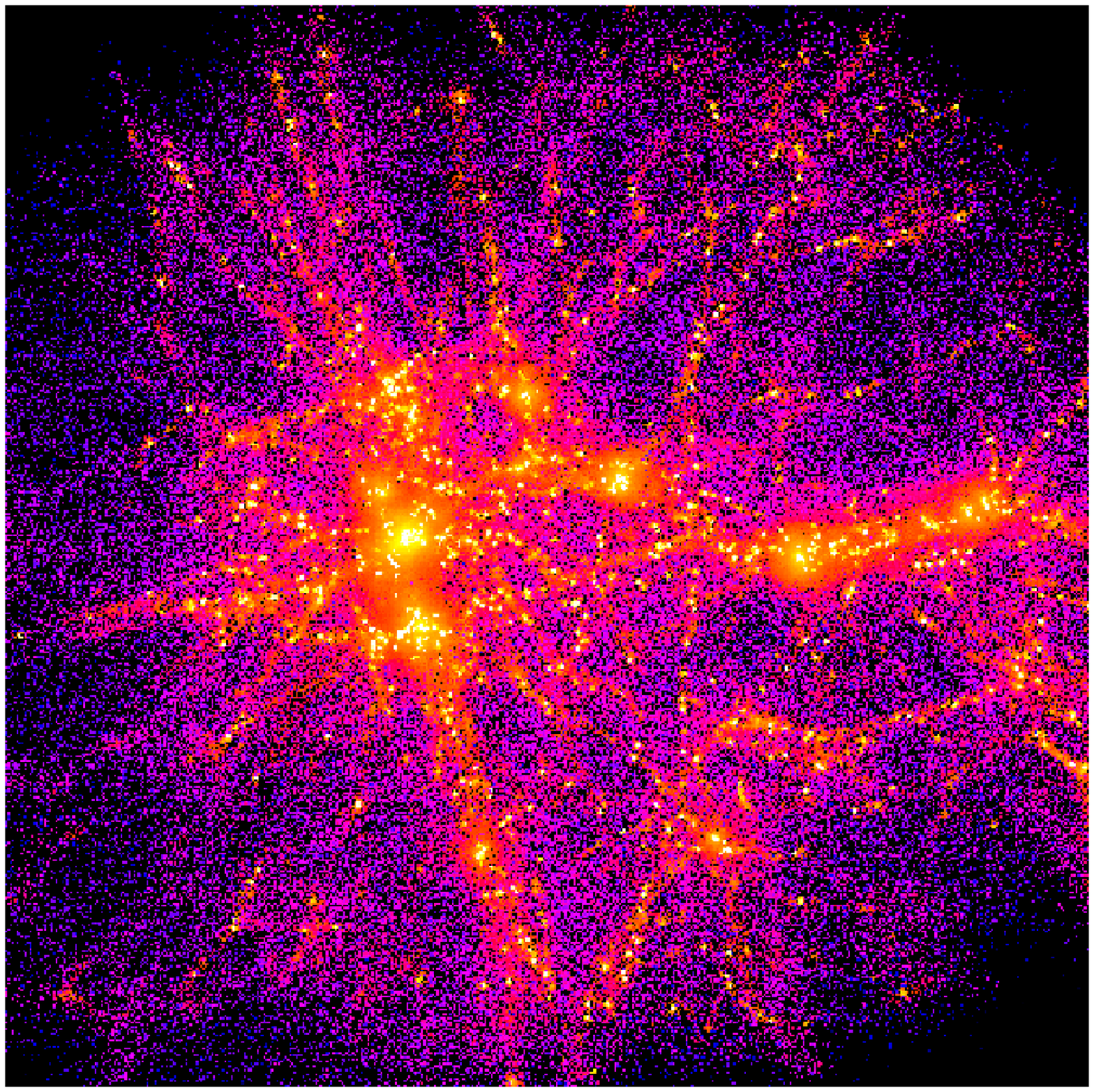,width=6.2cm} 
\psfig{file=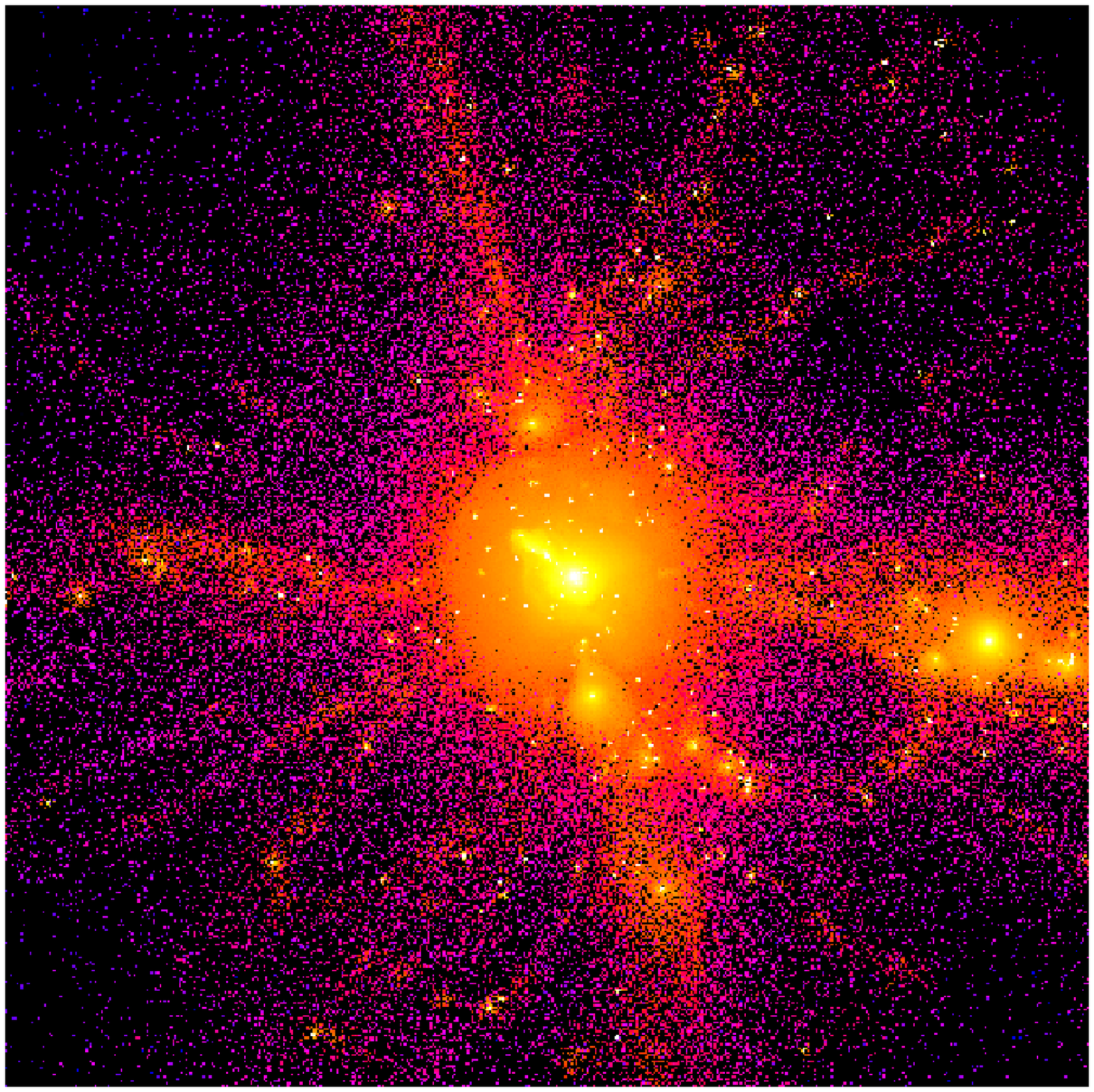,width=6.2cm} 
}
\hbox{
\psfig{file= 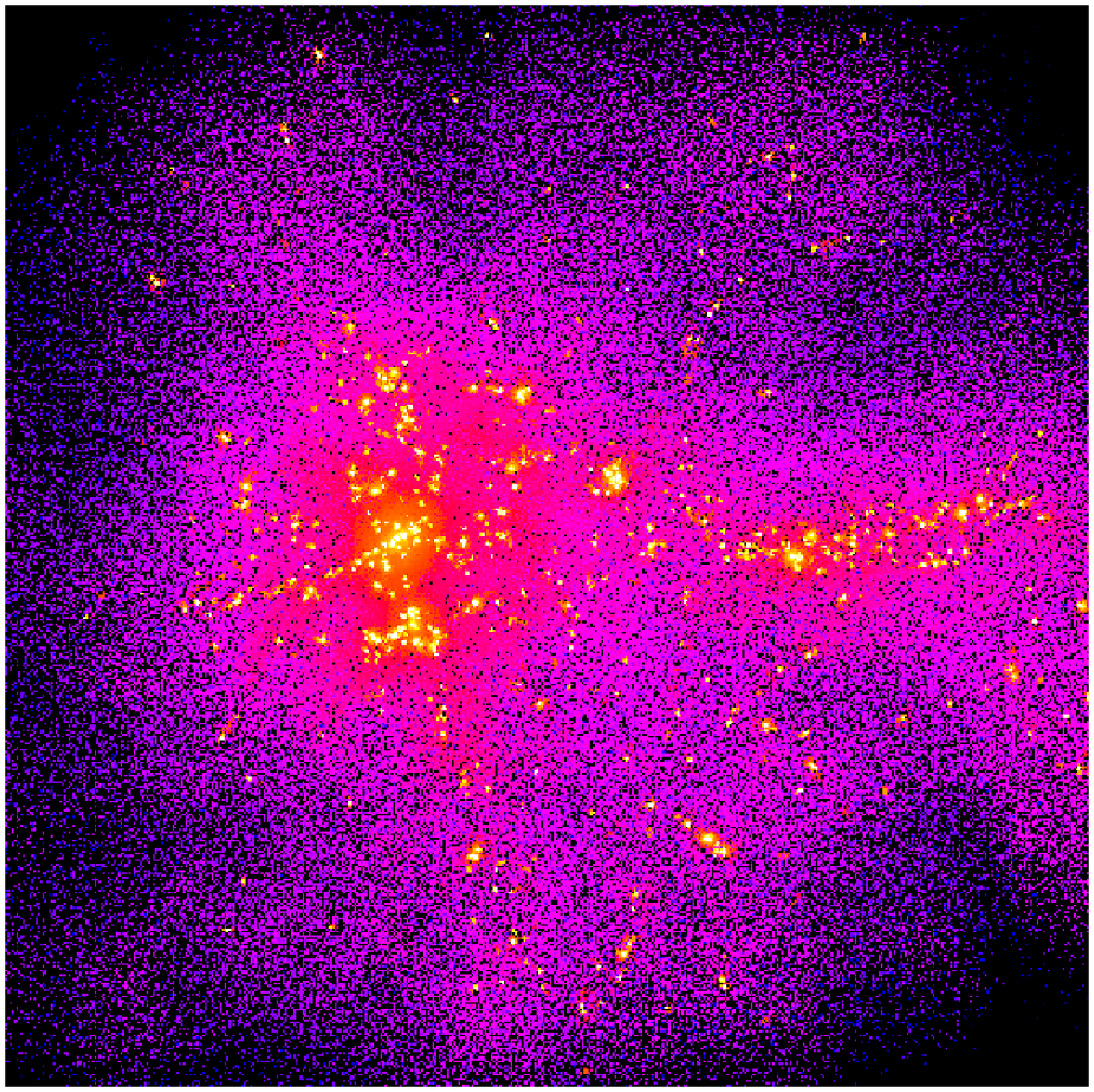,width=6.2cm} 
\psfig{file=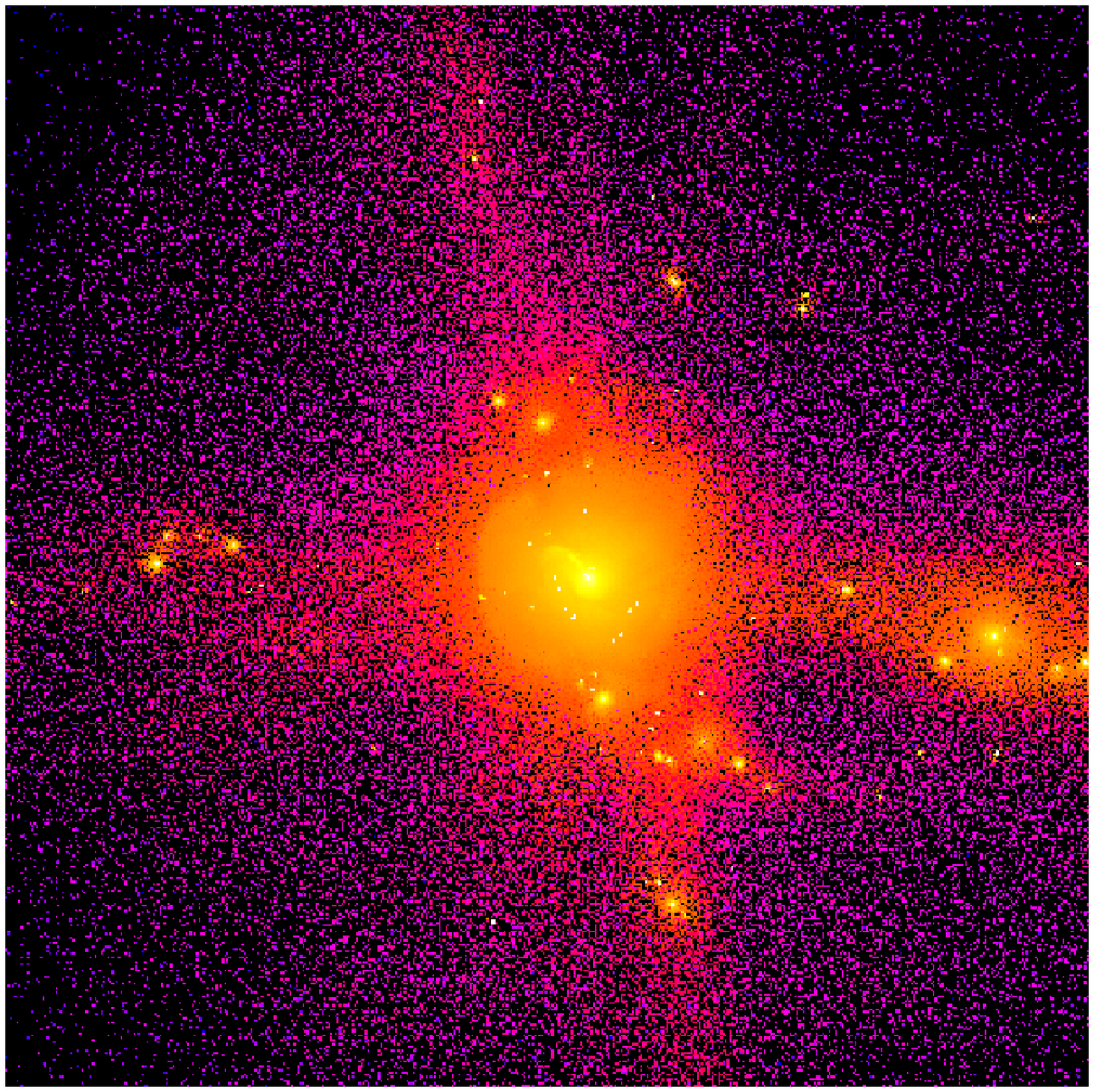,width=6.2cm} 
}
}}
\caption{The maps of the gas density for simulations of a Virgo--like
  cluster at $z=2$ and $z=0$ (left and right panels, respectively),
  including cooling, star formation and different forms of
  non--gravitational heating. Upper panels are for a run which
  includes galactic winds with a velocity of about 340$\vel$, the
  central panels is for galactic winds with velocity of about
  $830\vel$ and the bottom panels is for galactic winds as in the top
  panels, but also adding a pre--heating with an entropy threshold of
  100 keV cm$^2$ at $z=3$.  At $z=0$ the size of the box is of
  $11.7\hm$, while at $z=2$ is corresponds to $17.5\hm$ comoving (from
  \protect\citealt{2005MNRAS.361..233B}).}
\label{fi:csf} 
\end{figure*}

As an example, we show in Fig.~\ref{fi:csf} how the gas density of a
simulated cluster changes, both at $z=2$ (left panels) and at $z=0$
(right panels), when cooling and star formation are combined with
different forms of non--gravitational heating (from \citealt{2005MNRAS.361..233B}). The comparison of the top and
central panels shows the effect of increasing the kinetic energy
carried by galactic outflows by a factor of six. The stronger winds
are quite efficient in stopping star formation in the small halos,
which are washed out, and make the larger ones slightly puffier, while
preserving the general structure of the cosmic web surrounding the
Lagrangian cluster region. Comparing the top and the bottom panels
shows instead the effect of adding to galactic winds also the effect
of an entropy floor. Although the energy budget of the feedback
schemes of the central and bottom panels are quite comparable, the
effect of the gas distribution is radically different. Imposing an
entropy floor at $z=3$ with an impulsive heating generates a much
smoother gas density distribution, both at $z=2$ and at $z=0$. In this
case, the filamentary structure of the gas distribution is completely
erased, while only the largest halos are able to retain part of their
gas content. This demonstrates that a fixed amount of energy feedback
can provide largely different results on the ICM thermodynamical
properties, depending on the epoch and on the density at which it is
released.

\section{Scaling relations}
\label{scalings}
So far, we have qualitatively discussed how simple models of
pre--heating and radiative cooling can reproduce the observed
violation of self--similarity in the X--ray properties of galaxy
clusters. In this and in the following sections we will focus the
discussion on a more detailed comparison between simulation results
and observational data, and on the implications of this comparison on
our current understanding of the feedback mechanisms which regulate
star formation and the evolution of the galaxy population. As a
starting point for the comparison between observed and simulated
X--ray cluster properties, we describe how observable quantities are
computed from hydrodynamical simulations and how they compare to the
analogous quantities derived from observational data.

As for the X--ray luminosity, it is computed by summing the
contributions to the emissivity, $\epsilon_i$, carried by all the gas
elements (particles in a SPH run and cells in an Eulerian grid--based
run), $L_X=\sum_i\epsilon_i$, where the sum extends over all the gas
elements within the region where $L_X$ is computed. The contribution
from the $i$-th gas element is usually written as
\be
\epsilon_i\,=\,n_{{\rm e},i}n_{{\rm H},i}\Lambda(T_i,Z_i){\rm d}V_i\,,
\label{eq:emiss}
\ee
where $n_{{\rm e},i}$ and $n_{{\rm H},i}$ are the number densities of electrons
and of hydrogen atoms, respectively, associated to the $i$-th gas
element of given density $\rho_i$, temperature $T_i$ and metallicity
$Z_i$. Furthermore, $\Lambda(T,Z)$ is the temperature-- and
metallicity--dependent cooling function (e.g., 
\citealt{1993ApJS...88..253S}) computed within a given energy band,
while ${\rm d}V_i=m_i/\rho_i$ is the volume of the $i$-th gas element,
having mass $m_i$.

As for the temperature, different proxies to its X--ray observational
definition have been proposed in the literature, which differ from each
other in the expression for the weight assigned to each gas element. In
general, the ICM temperature can be written as
\be
T\,=\,{\sum\limits_i^{} w_i T_i\over \sum\limits_i^{} w_i}\,,
\label{eq:tgen}
\ee
where $T_i$ is the temperature of the $i$-gas element, which
contributes with the weight $w_i$. The mass--weighted definition of
temperature, $T_{\rm mw}$, is recovered for $w_i=m_i$ ($m_i$: mass of the
$i$-th gas element), which also coincides with the electron
temperature $T_{\rm e}$ for a fully ionised plasma. A more
observation--oriented estimate of the ICM temperature is provided by the
emission--weighted definition, $T_{\rm ew}$, which is obtained for
$w_i=\epsilon_i$ (e.g.,  \citealt{1996ApJ...469..494E}). The idea underlying this definition
is that each gas element should contribute to the overall spectrum
according to its emissivity.

\citet{2004MNRAS.354...10M} pointed out that the
thermal complexity of the ICM is such that the overall spectrum is
given by the superposition of several single--temperature spectra,
each one associated to one thermal phase. In principle, the
superposition of several single--temperature  spectra
cannot be described by a single--temperature spectrum. However, when
fitting it to a single--temperature model in a typical finite energy
band, where X--ray telescopes are sensitive, the cooler gas phases are
relatively more important in providing the high--energy cut--off of
the spectrum and, therefore, in determining the temperature resulting
from the spectral fit. In order to account for this effect, \citet{2004MNRAS.354...10M} introduced a spectroscopic--like
temperature, $T_{\rm sl}$, which is recovered from Eq.~\ref{eq:tgen} by
using the weight $w_i=\rho_i m_i T^{\alpha-3/2}$. By using
$\alpha=0.75$, this expression for $T_{\rm sl}$ was shown to reproduce
within few percent the temperature obtained from the spectroscopic
fit, at least for clusters with temperature above 2--3 keV. More
complex fitting expressions have been provided by \citet{2006ApJ...640..710V}, who generalised the spectroscopic--like
temperature to the cases of lower temperature and arbitrary
metallicity.

\subsection{The luminosity--temperature relation}
The $L_X$--$T$ relation represented the first observational evidence
against the self--similar model. This relation has been shown by
several independent analyses to have a slope, $L_X\propto T^\alpha$,
with $\alpha\simeq 3$ for $T\gtrsim 2$ keV (e.g., \citealt{1997MNRAS.292..419W}), with indications for a flattening to
$\alpha\lesssim 2.5$ for the most massive systems
\citep{1998MNRAS.297L..57A}. The scatter in this relation is largely
contributed by the cool--core emission, so that it significantly
decreases when excising the cores \citep{1998ApJ...504...27M} or
removing cool--core systems \citep{1999MNRAS.305..631A}. A change of
behaviour is also observed at the scales of groups, $T\lesssim 2$ keV,
which generally displays a very large scatter
\citep{2004MNRAS.350.1511O}.

Hydrodynamical simulations by \citet{2001ApJ...555..597B} and by  \citet{2002MNRAS.336..409B} demonstrated that simple
pre--heating models, based on the injection of entropy at relatively
high redshift, can reproduce the observed slope of the $L_X$--$T$
relation. \citet{2002ApJ...579...23D} showed that a
similar result can also be achieved with simulations including cooling
only, the price to be paid being a large overcooling. \citet{2002MNRAS.336..527M} and \citet{2003MNRAS.342.1025T} demonstrated that combining cooling
with pre--heating models can eventually decrease the total amount of
stars to an acceptable level, while still providing a slope of the
$L_X$--$T$ relation close to the observed one.

\begin{figure}
\hbox{
\psfig{figure=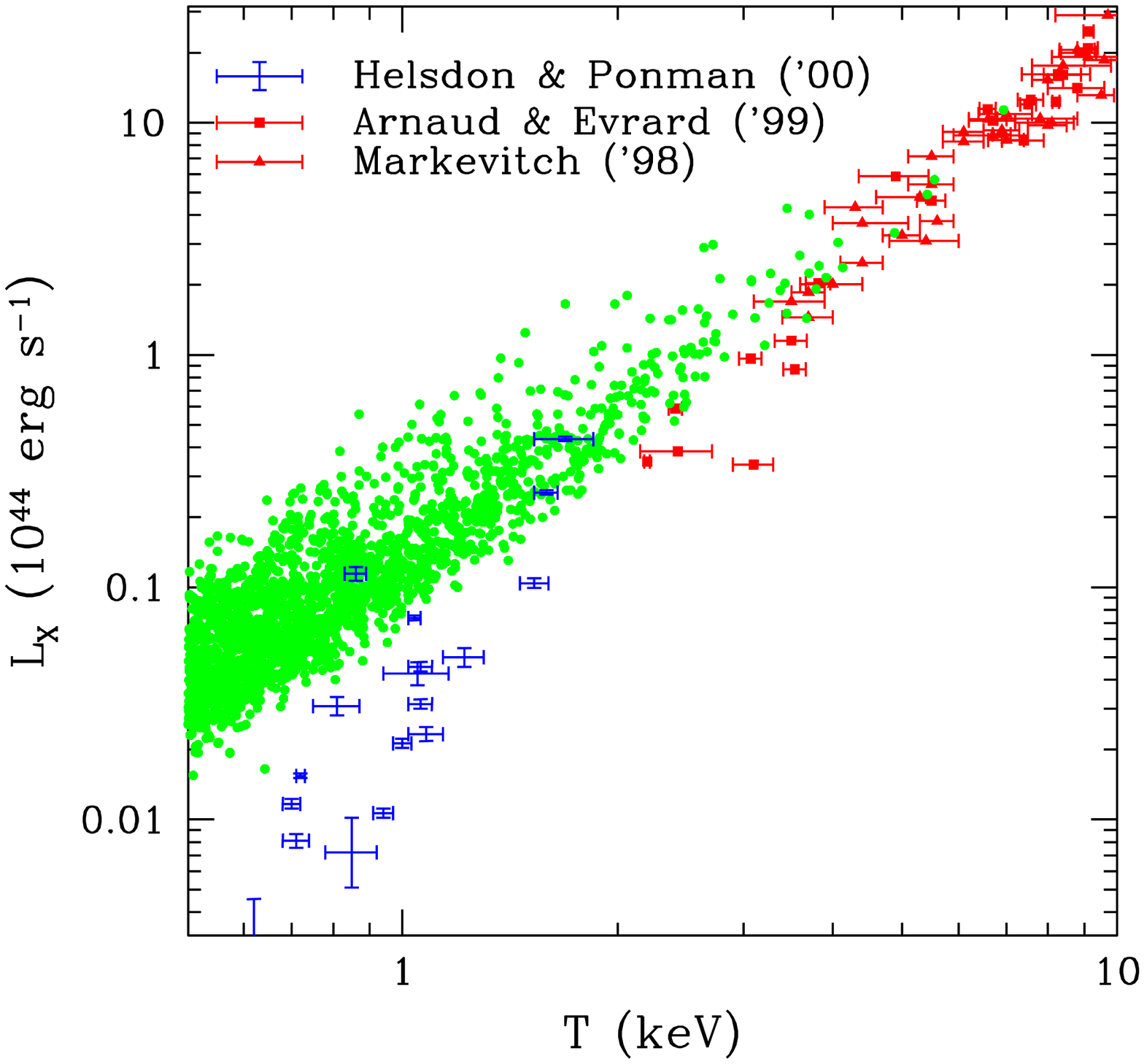,height=5.5cm}
\psfig{figure=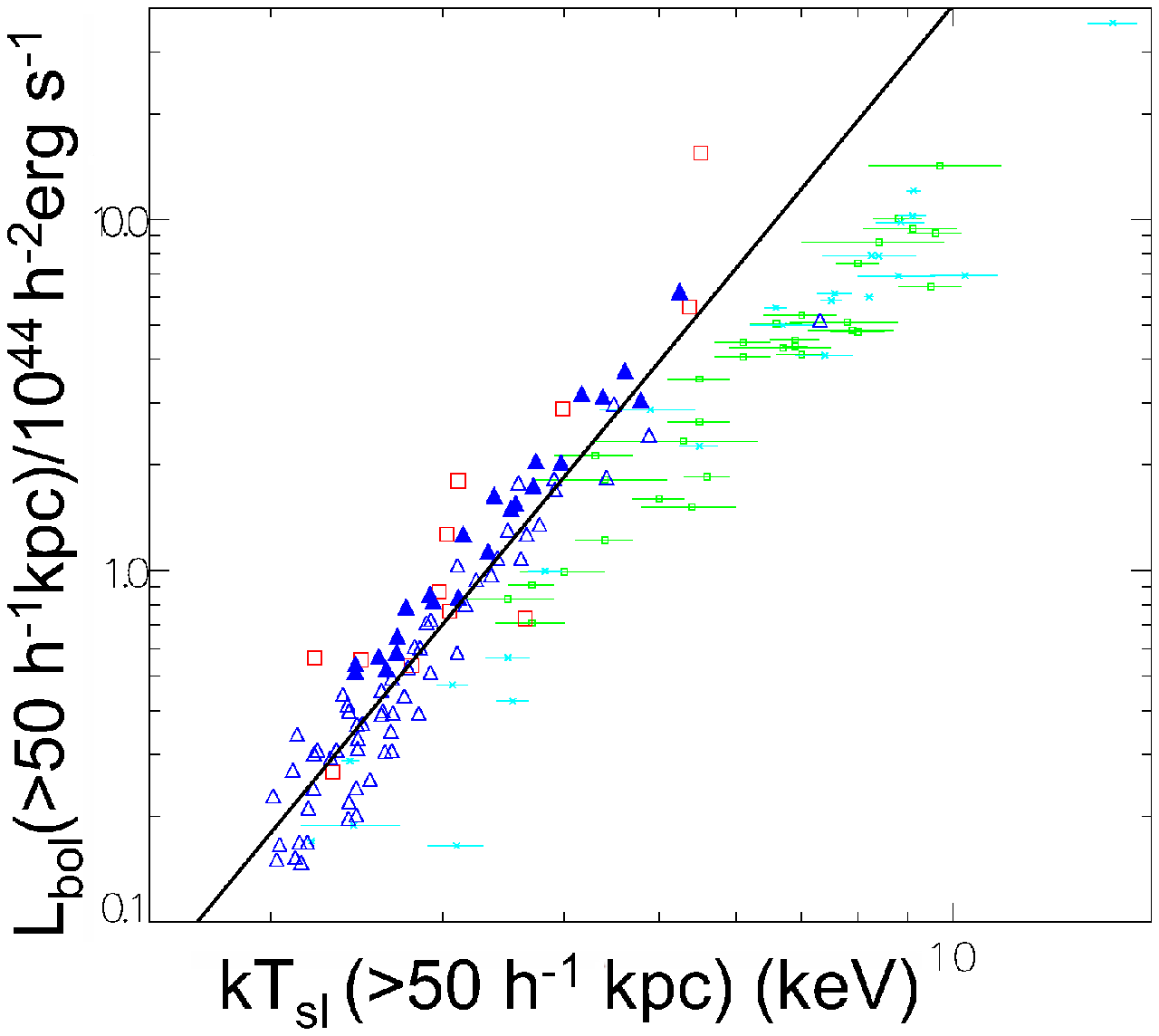,height=5.5cm}
}
\caption{Left panel: The relation between bolometric X--ray luminosity
  and emission--weighted temperature for simulations by \protect\citet{2004MNRAS.348.1078B}, including cooling, star
  formation and feedback in the form of galactic winds (green circles),
  compared with observational data for clusters
  \protect\citep{1999MNRAS.305..631A,1998ApJ...504...27M} and groups
  \protect\citep{2000MNRAS.315..356H}. Right panel: the relation
  between X--ray luminosity, estimated outside the core regions, and
  the spectroscopic--like temperature, for simulations which include
  cooling, star formation and a form of ``targeted'' feedback (from
  \protect\citealt{2007MNRAS.377..317K}). Observational data
  for clusters are the same as in the left panel.}
\label{fi:lt}
\end{figure}

In Fig.~\ref{fi:lt} we show the comparison between observations and
simulations in which the non--gravitational gas heating assumes the
energy budget made available by the SN explosions, whose rate is
computed from the simulated star formation rate. The left panel shows
the results from the SPH {\tt GADGET} simulations by \citet{2004MNRAS.348.1078B}. These simulations included the same
model of kinetic feedback used in the simulations shown in the top
panel of Fig.~\ref{fi:csf}. This feedback model was shown by
\citet{2003MNRAS.339..312S} to be quite
successful in producing a cosmic star formation history similar to the
observed one. Quite apparently, these simulations provided a
reasonable relation at the scale of clusters, $T\gtrsim 3$ keV, while
failing to produce slope and scatter at the scale of groups.

The right panel show the results by \citet{2007MNRAS.377..317K}, which these authors plotted only
for systems with $T>2$ keV. Kay et al. also used SPH simulations based
on the {\tt GADGET} code, but with a different feedback scheme. In
this scheme, energy made available by SN explosions is assigned in a
``targeted'' way to suitably chosen gas particles, which surround the
star--forming regions, so that their entropy is raised in such a way
to prevent them from cooling. Therefore, while the amount of energy is
self--consistently computed from star formation, the way in which it
is thermalised to the gas is suitably tuned. These simulations predict
a too high normalisation of the $L_X$--$T$ relation. This result is
interpreted by \citet{2007MNRAS.377..317K} as due to the
fact that their simulations produce too low temperatures for clusters
of a given mass, as a consequence of the incorrect cool--core
structure.

Besides the slope of the local $L_X$--$T$ relation, also its
  evolution carries information about the thermodynamical history of
  the ICM. Thanks to the increasing statistics of distant clusters
  observed in the last years with Chandra and XMM--Newton, a number of
  authors analysed this evolution out to the highest redshifts,
  $z\simeq 1.3$, where clusters have been detected so far. These
  analyses generally indicate that the amplitude of the $L_X$--$T$
  relation has a positive evolution out to $z\simeq 0.5$--0.6
  (e.g. \citealt{2002A&A...389....1A,2004A&A...420..853L,2005ApJ...633..781K}), with hints for a possible inversion of
  this trend at higher redshift (e.g. \citealt{2004A&A...417...13E,2006MNRAS.365..509M,2007A&A...472..739B}). In the attempt of interpreting
  these results, \citet{2005RvMP...77..207V} showed that
  radiative cooling, combined with a modest amount of pre--heating
  with extra entropy, predicts an evolution of the $L_X$--$T$ relation
  slower than that of the self--similar model, also with an inversion
  of the trend at high redshift. Although in qualitative agreement
  with observations, these models have however difficulties in
  following the observed positive evolution at $z\lesssim 0.5$ (e.g.,
  see Fig.~14 by \citealt{2006MNRAS.365..509M}).

\begin{figure}
\hbox{
\psfig{figure=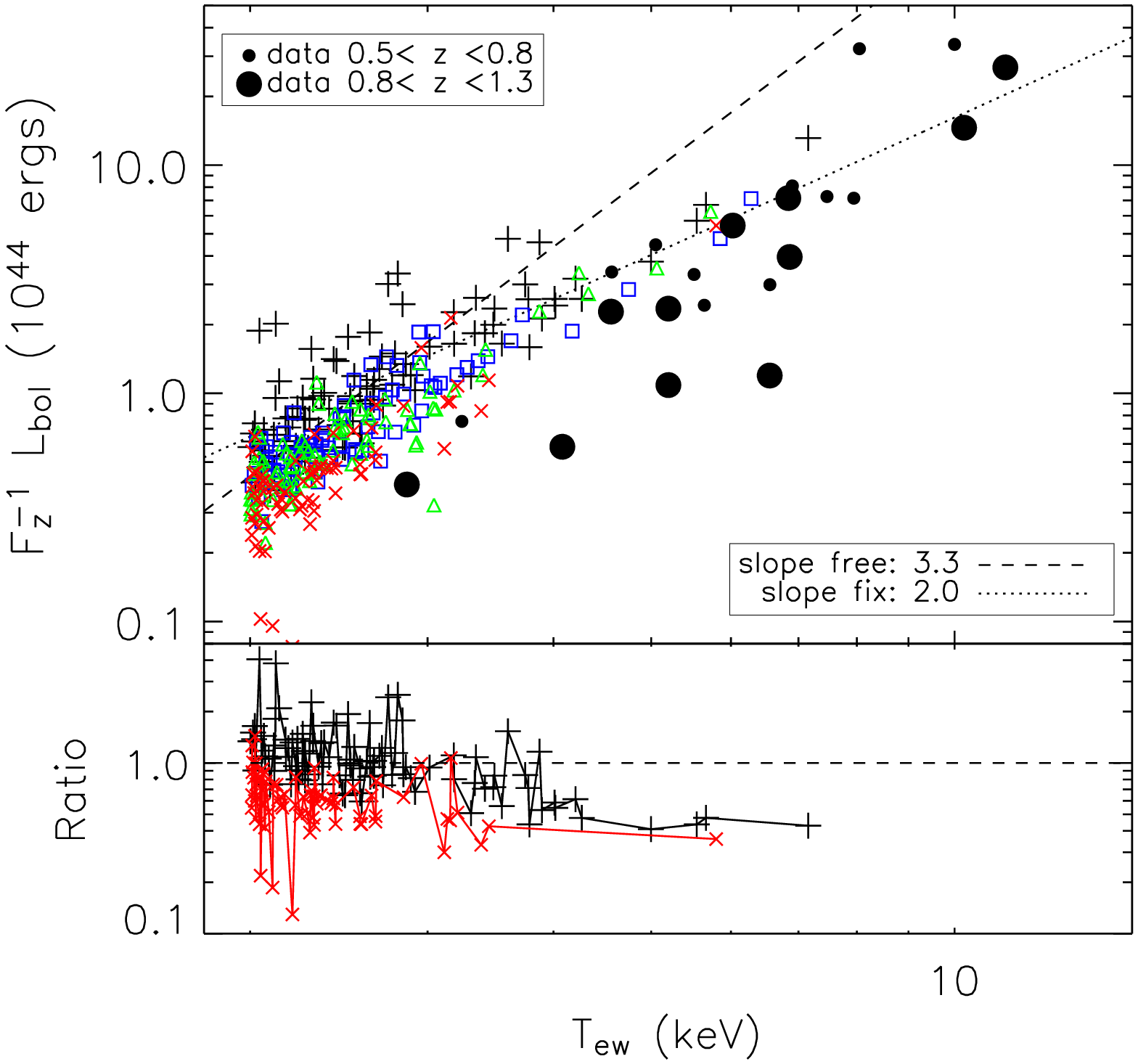,height=6.cm}
\psfig{figure=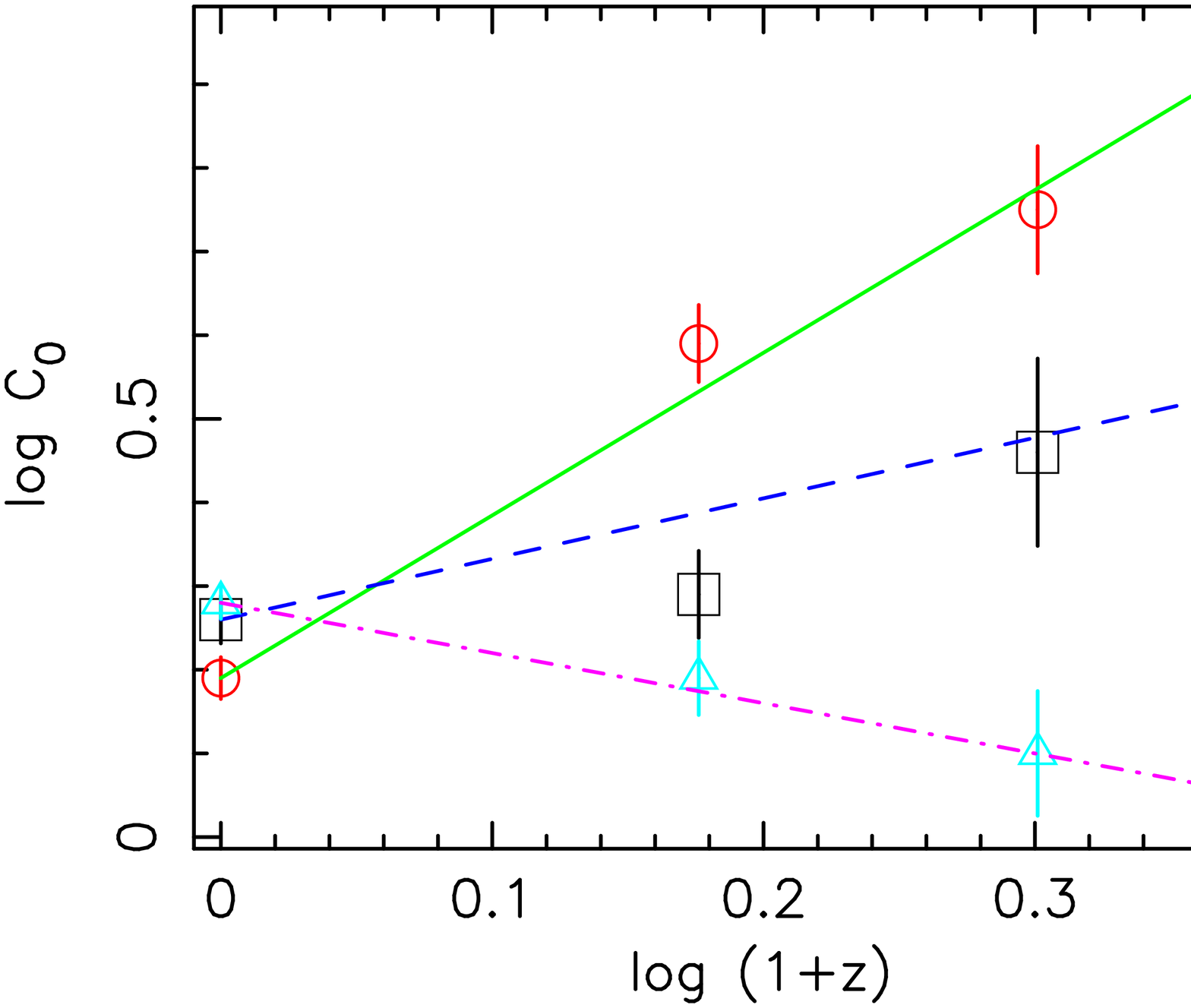,height=5.cm}
}
\caption{Left panel: A comparison of the evolution of the $L_X$--$T$
  relation for the simulated clusters by \protect\citet{2004MNRAS.348.1078B} and the Chandra data
  analysed by \protect\citet{2004A&A...417...13E}
  (from \protect\citealt{2004MNRAS.354..111E}). The
  ``plus'' symbols are the relation at $z=0$, with the dotted and the
  dashed lines providing the best--fit relations when the slope is
  kept fixed at 2 and is left free, respectively. The squares, the
  triangles and the crosses show the simulation results at $z=0.5$,
  0.7 and 1, respectively. The small and the big dots show the
  observational data at $0.5<z<0.8$ and at $z>0.8$, respectively. The
  lower panel shows the ratio between the measured luminosities at
  $z=0$ (``plus'' symbols) and at $z=1$ (crosses) and the
  corresponding best--fitting power laws.  Right panel: normalisation
  of the $L_X$--$T$ relation as a function of redshift, for the
  simulated clusters by \protect\citet{2006ApJ...649..640M} in the radiative run with
  no feedback (solid line), in a run including an impulsive
  pre--heating at $z=4.5$ (dashed line), and in a run including a form
  of ``targeted'' feedback (dot--dashed line).}
\label{fi:lt_ev}
\end{figure}

  As for the comparison with numerical predictions, \citet{2004MNRAS.354..111E} analysed the evolution of the
  $L_X$--$T$ relation from the radiative simulation by \citet{2004MNRAS.348.1078B}, which includes the effect of SN
  feedback in the form of galactic winds. As a result, they found that
  the normalisation of this relation for the simulated clusters at
  high redshift is higher than for real clusters (see left panel of
  Fig.~\ref{fi:lt_ev}). \citet{2006ApJ...649..640M} analysed three sets of simulated
  clusters, based on radiative cooling only, on gas pre--heating at
  high redshift and on the same ``targeted'' SN feedback model used by
  \citet{2007MNRAS.377..317K}.  As shown in the right
  panel of Fig.~\ref{fi:lt_ev}, they found that these three models
  predicts rather different evolutions of the $L_X$--$T$ relation,
  thus confirming it to be a sensitive test for the thermodynamical
  history of the ICM. However, so far none of the numerical models
  proposed is able to account for the observational indication for an
  inversion of the $L_X$--$T$ evolution at $z\simeq 0.5$. This
  represents still an open issue, whose implications for a realistic
  modelling of the ICM physics are still to be understood.

\subsection{The mass--temperature relation}
The relation between total collapsed mass and temperature has received
much consideration both from the observational and the theoretical
side, in view of its application for the use of galaxy clusters as
tools to measure cosmological parameters (e.g., 
\citealt{2005RvMP...77..207V,2006astro.ph..5575B}). The
relation between ICM temperature and total mass should be primarily
dictated by the condition of hydrostatic equilibrium. For this reason,
the expectation is that this relation should have a rather small
scatter and be insensitive to the details of the heating/cooling
processes. For a spherically symmetric system, the condition of
hydrostatic equilibrium translates into the mass estimator
(e.g. \citealt{1988xrec.book.....S})
\be
M(<r)\,=\,-{r {\rm k}_{\rm B}T(r)\over G \mu m_{\rm p}}\left[{{\rm d}\,\ln \rho_{\rm gas}(r)\over {\rm d}\,\ln r} +
{{\rm d}\,\ln T(r)\over {\rm d}\,\ln r}\right]\,.
\label{eq:he}
\ee
Here $M(<r)$ is the total mass within the cluster-centric distance
$r$, while $T(r)$ is the temperature measured at $r$. As shown in the
previous section, the processes of heating/cooling are expected to
modify the gas thermodynamics only in the central cluster regions,
while the bulk of the ICM is dominated by gravitational
processes. This implies that, in principle, the total mass estimate
from Eq.~\ref{eq:he} should be rather stable. However, since the
$X$--ray spectroscopic temperature is sensitive to the thermal
complexity of the ICM in the central regions (e.g., \citealt{2004MNRAS.354...10M}), a change
of temperature and density profile in these regions may translate into
a sizable effect in the mass--temperature relation provided by
Eq.~\ref{eq:he}.

Eq.~\ref{eq:he} has been often applied in the literature by modelling
the gas density profile with a $\beta$--model
\citep{1976A&A....49..137C},
\be
\rho_{\rm gas}(r)\,=\,{\rho_0 \over \left[1+(r/r_{\rm c})^2\right]^{3\beta /2}}\,,
\label{eq:betam}
\ee
where $r_{\rm c}$ is the core radius, and assuming either isothermal gas
or a polytropic
gas, $\rho_{\rm gas}\propto T^{\gamma -1}$, to account for the presence of
temperature gradients. In this case, Eq.~\ref{eq:he} can
be recast in the form
\be 
M(<r)\,\simeq \,1.1\times 10^{14}\beta \gamma \,{T(r)\over
  {\rm keV}}\,{r\over \hm}\,{(r/r_{\rm c})^2 \over 1+(r/r_{\rm c})^2}\,.
\label{eq:hebg}
\ee

\begin{figure}
\centerline{
\psfig{figure=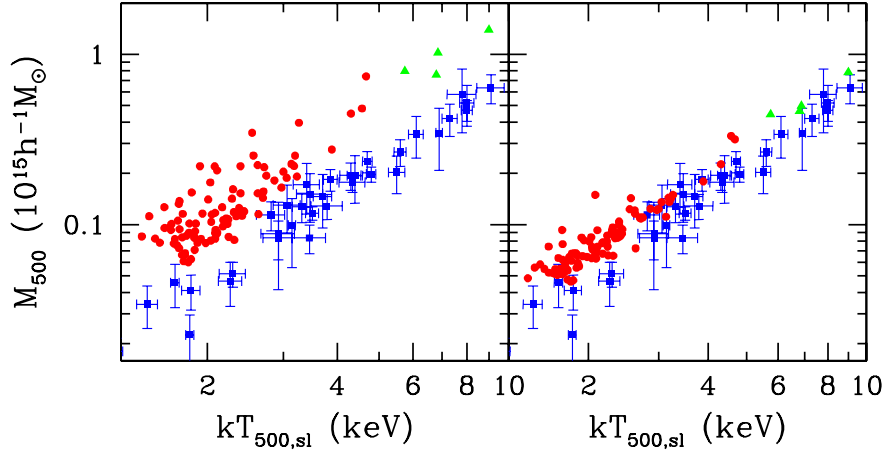,width=13cm}
}
\vspace{-5.7truecm}
\caption{The relation between mass and spectroscopic--like temperature
  within $r_{500}$. Red circles and green triangles are for
  simulations, which include cooling, star formation and feedback from
  galactic winds, while squares with error bars are the observational
  data by \protect\citet{2001A&A...368..749F}. Left panel: $M_{500}$
  exactly computed by summing the mass of all the particles within
  $r_{500}$. Right panel: $M_{500}$ estimated as in the observational
  data, by using the equation of hydrostatic equilibrium for a
  polytropic $\beta$--model. From \protect\citet{2005ApJ...618L...1R}.}
\label{fi:mt_rasia}
\end{figure}

Based on the analysis of non--radiative cluster simulations, \citet{1996A&A...305..756S} and \citet{1996ApJ...469..494E} argued that the X--ray temperature
provides a rather precise determination of the cluster mass, with an
intrinsic scatter of only about 15 per cent at $r_{500}$. \citet{2001A&A...368..749F} applied Eq.~\ref{eq:hebg} to
ROSAT imaging and ASCA spectroscopic data (see also \citealt{2000ApJ...532..694N}). They found that the resulting
$M$--$T$ relation has a normalisation about 40 per cent lower than
that from the simulations by \citet{1996ApJ...469..494E}. Although introducing the effect of
cooling, star formation and SN feedback provides a 20 per cent lower
normalisation, this was not yet enough to recover agreement with
observations. Independent analyses
\citep{2002MNRAS.336..527M,2004MNRAS.348.1078B} showed that applying to
simulated clusters Eq.~\ref{eq:hebg}, which is used to estimate
masses of real clusters, leads to a mass underestimate of about 20 per
cent. This bias in the mass estimate is enough to bring the simulated
and the observed $M$--$T$ relations into reasonable agreement. These
analyses were still based on the emission--weighted temperature in the
simulation analysis.

\citet{2005ApJ...618L...1R} showed that using the
spectroscopic--like definition, $T_{\rm sl}$, leads to a mass
underestimate of up to $\sim 30$ per cent with respect to the true
cluster mass. This result is shown in Fig.~\ref{fi:mt_rasia}. The
left panel reports the relation between $T_{\rm sl}$ for simulated
clusters and the true total cluster mass, also compared with the
observed $M$--$T$ relation by \citet{2001A&A...368..749F}. Quite apparently, the relation from
simulations lies well above the observational one. However, this
difference is much reduced in the right panel, where the masses of the
simulated clusters are computed by applying Eq.~\ref{eq:hebg}. The
reason for this difference between ``true'' and ``recovered'' masses
is partly due to the violation of hydrostatic equilibrium, associated
to subsonic gas bulk motions (e.g., 
\citealt{2004MNRAS.351..237R,2007ApJ...655...98N})
and partly to the poor fit provided by the $\beta$--model (e.g.,
\citealt{2003MNRAS.346..731A}) when extended to large
radii.  It is also interesting to note that the mass estimator of
Eq.~\ref{eq:hebg} also under-predicts the intrinsic scatter of the
$M$--$T$ relation from the simulations.  This is due to the fact that
the equation of hydrostatic equilibrium imposes a strong correlation
between ICM temperature and total mass. Any scatter is then associated
to a cluster-by-cluster variation of the parameters $\beta$ and
$\gamma$, which may not be fully representative of the diversity of
the ICM structure among different objects.

\begin{figure}
\centerline{
\hbox{
\psfig{figure=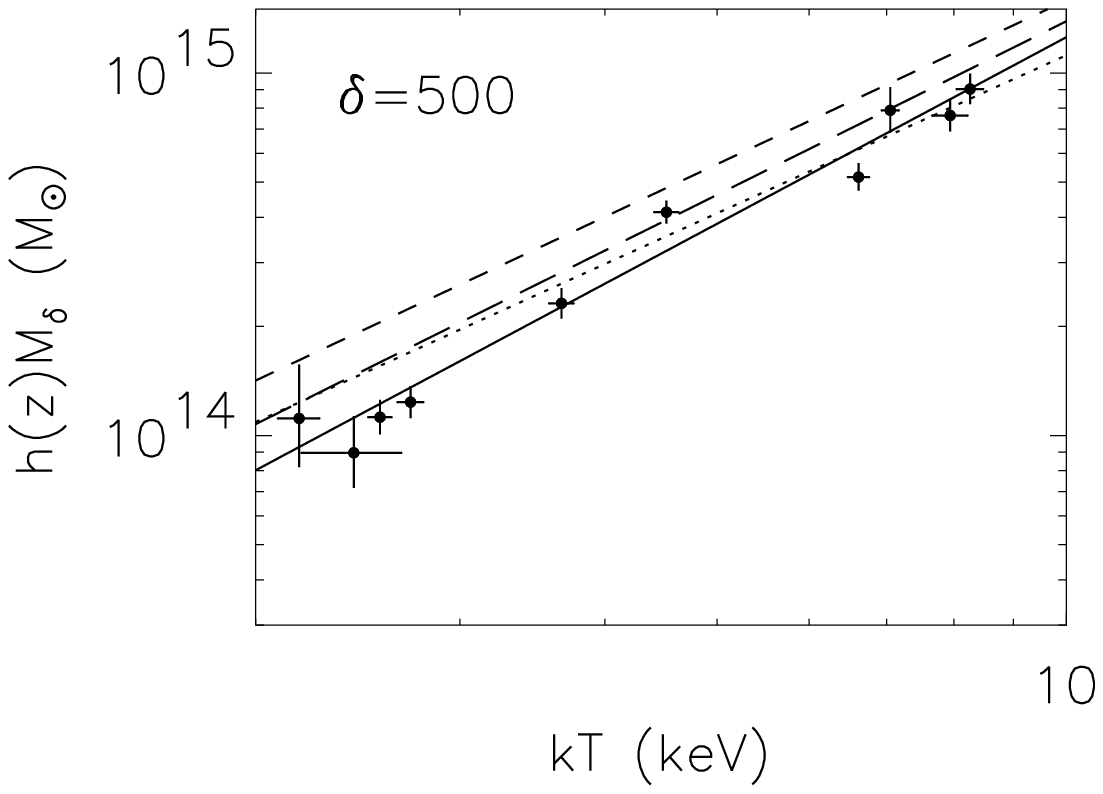,height=5.5cm}
\psfig{figure=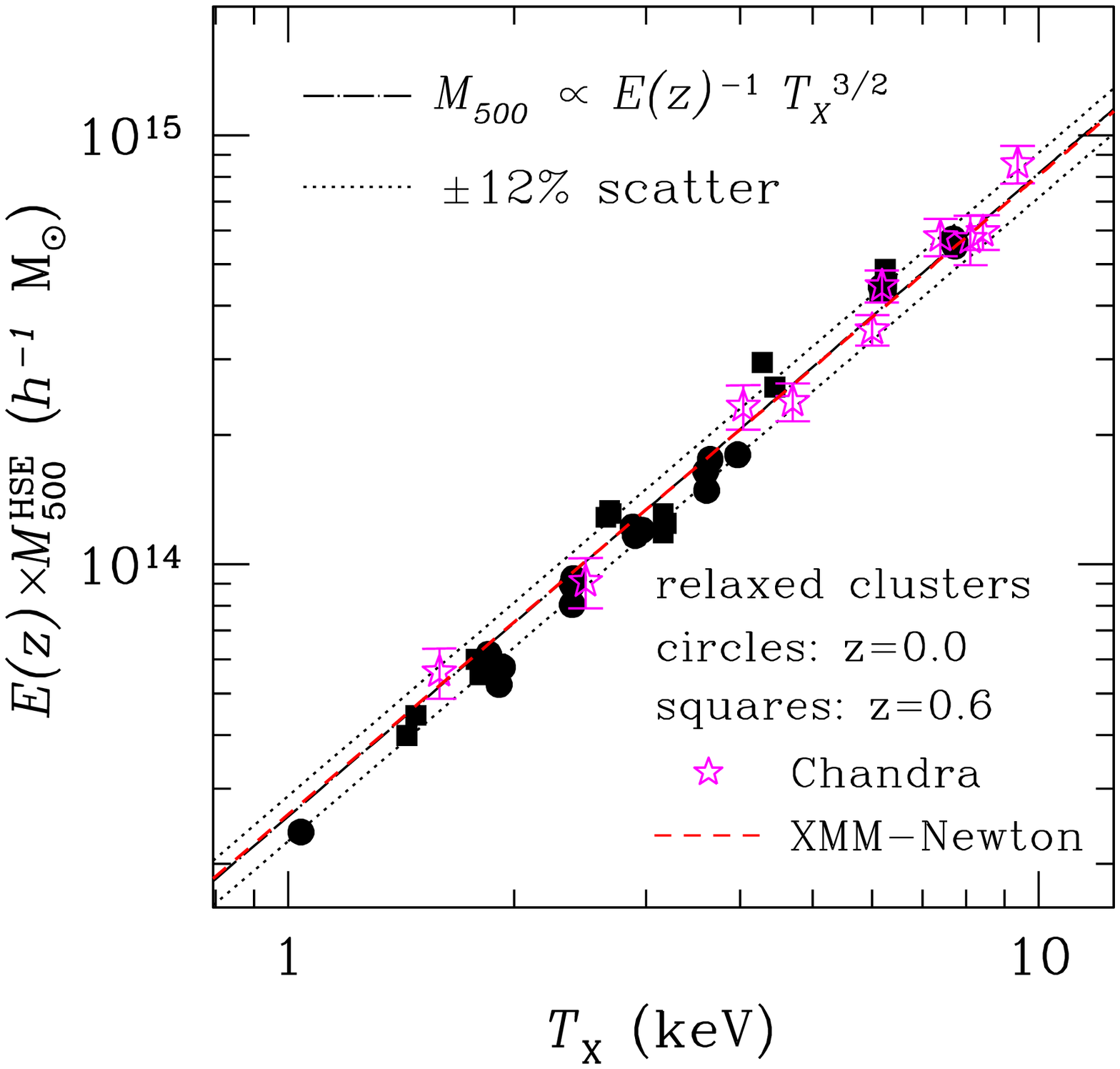,width=6.5cm}
}}
\caption{Left panel: the observed mass--temperature relation at
  $r_{500}$ for nearby clusters observed with XMM--Newton, compared
  with $M$--$T$ relations from simulations (from \protect\citealt{2005A&A...441..893A}). The solid line is the best
  fit to observations, while the dotted line is the best fit only to a
  subsample of hot clusters. The short-dashed line is the relation
  from the non--radiative simulations by \protect\citet{1996ApJ...469..494E}, while the long--dashed
  line is the relation from the radiative simulations by \protect\citet{2004MNRAS.348.1078B}. Right panel: the $M$--$T$
  relation at $r_{500}$ from the radiative simulations by \protect\citet{2007astro.ph..3661N}; circles: $z=0$; squares:
  $z=0.6$, which include cooling, star--formation and an inefficient
  form of SN feedback, compared to Chandra observations for a set of
  nearby relaxed clusters by \protect\citet{2005ApJ...628..655V}; stars with
  error bars. Masses of simulated clusters have been computed by using
  the same estimator, based on the hydrostatic equilibrium, applied to
  the Chandra data. Masses of both simulated and observed clusters
  have been rescaled with redshift according to the evolution
  predicted by the self--similar model.}
\label{fi:mt}
\end{figure}

Using XMM--Newton and Chandra data, different authors (e.g., \citealt{2005A&A...441..893A,2005ApJ...628..655V}) applied the equation of hydrostatic
equilibrium by avoiding the assumption of a simple beta--model for the
gas density profile. As a result, the observed and the simulated
$M$--$T$ relations turned out to agree with each other, especially
when simulated clusters are analysed in the same way as real clusters
(e.g.,  \citealt{2006MNRAS.369.2013R,2007ApJ...655...98N}). The left panel of Fig.~\ref{fi:mt}
shows the comparison between the observational results by \citet{2005A&A...441..893A} and simulations. Here the
discrepancy with respect to the non--radiative runs by \citet{1996ApJ...469..494E} is alleviated when including the
effect of star formation and SN feedback
\citep{2004MNRAS.348.1078B}. In a similar way, the right panel shows
the comparison between the observed $M$--$T$ relation by \citet{2005ApJ...628..655V} and the simulations by \citet{2007astro.ph..3661N} who computed cluster masses by using
the same method applied to the Chandra data. This plot further
demonstrates the good level of agreement between simulated and
observed $M$--$T$ relation, once the 15--20 percent violation of
hydrostatic equilibrium is taken into account.

\subsection{The mass--''pressure'' relation}
To first approximation, the ICM can be represented by a smooth gas
distribution in pressure equilibrium within the cluster potential
well. Therefore, the expectation is that gas pressure should be the
quantity that is more directly correlated to the total collapsed
mass. For this reason, any pressure--related observational quantity
should provide a robust minimum--scatter proxy to the cluster
mass. One such observable is the Comptonisation parameter, measured
through the Sunyaev-Zeldovich effect (SZE; e.g., 
\citealt{2002ARA&A..40..643C}, for a review), which is proportional to
the ICM pressure integrated along the line of sight. Observations
  of the SZ effect are now reaching a high enough quality for extended
  sets of clusters, to allow correlating the SZ signal with X--ray
  observable quantities. For instance, \citet{2007arXiv0708.0815B} analysed a set of 38 massive
  clusters in the redshift range $0.14\le z\le 0.89$, for which both
  SZE imaging from the OVRO/BIMA interferometric array and Chandra
  X--ray observations are available. As a result, they found that the
  slope and the evolution of the scalings of the Comptonisation
  parameter with gas mass, total mass and X--ray temperature are all
  in agreement with the prediction of the self--similar model.

\begin{figure}
\centerline{
\hbox{
\psfig{figure=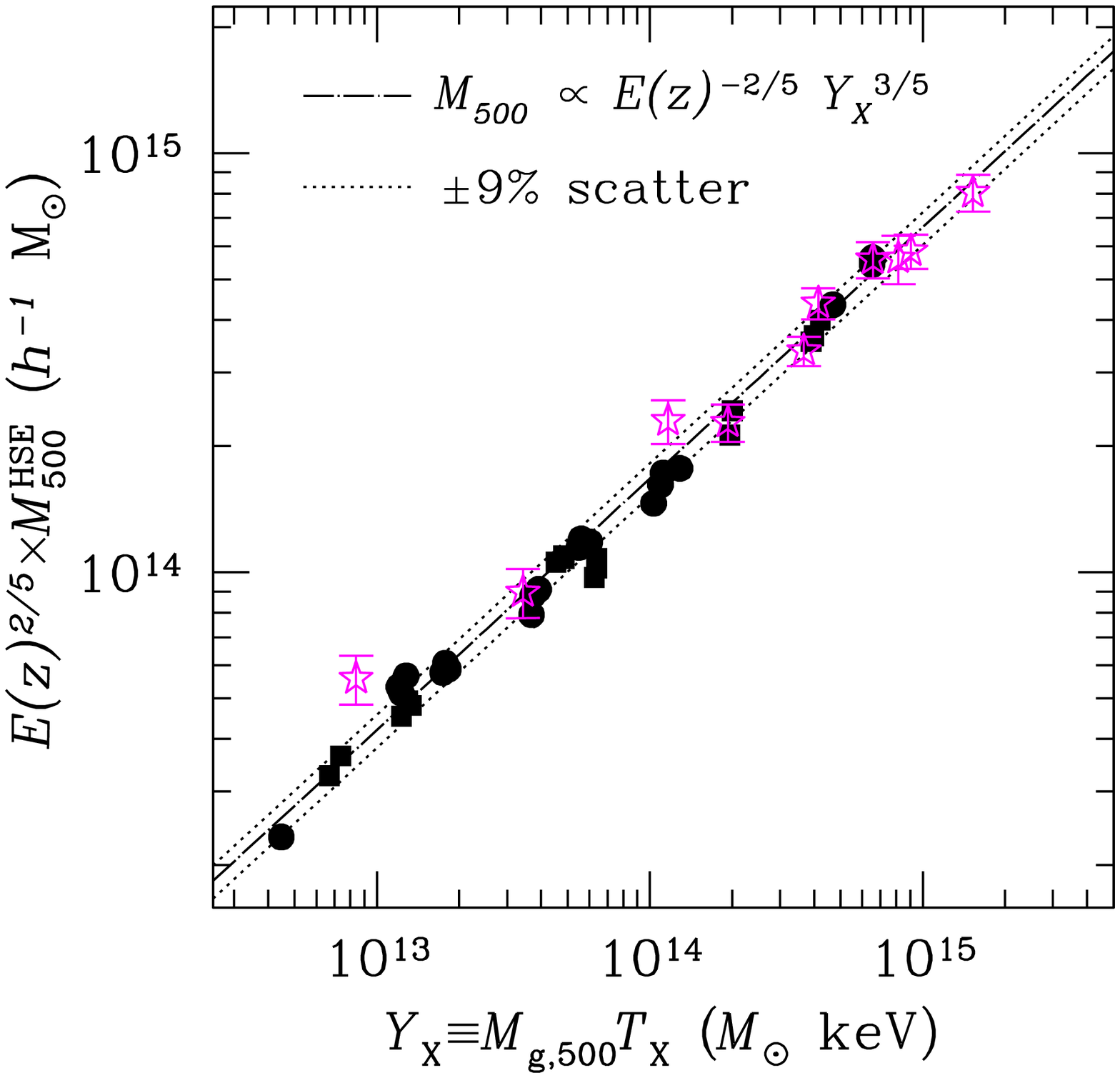,width=6.5cm}
\psfig{figure=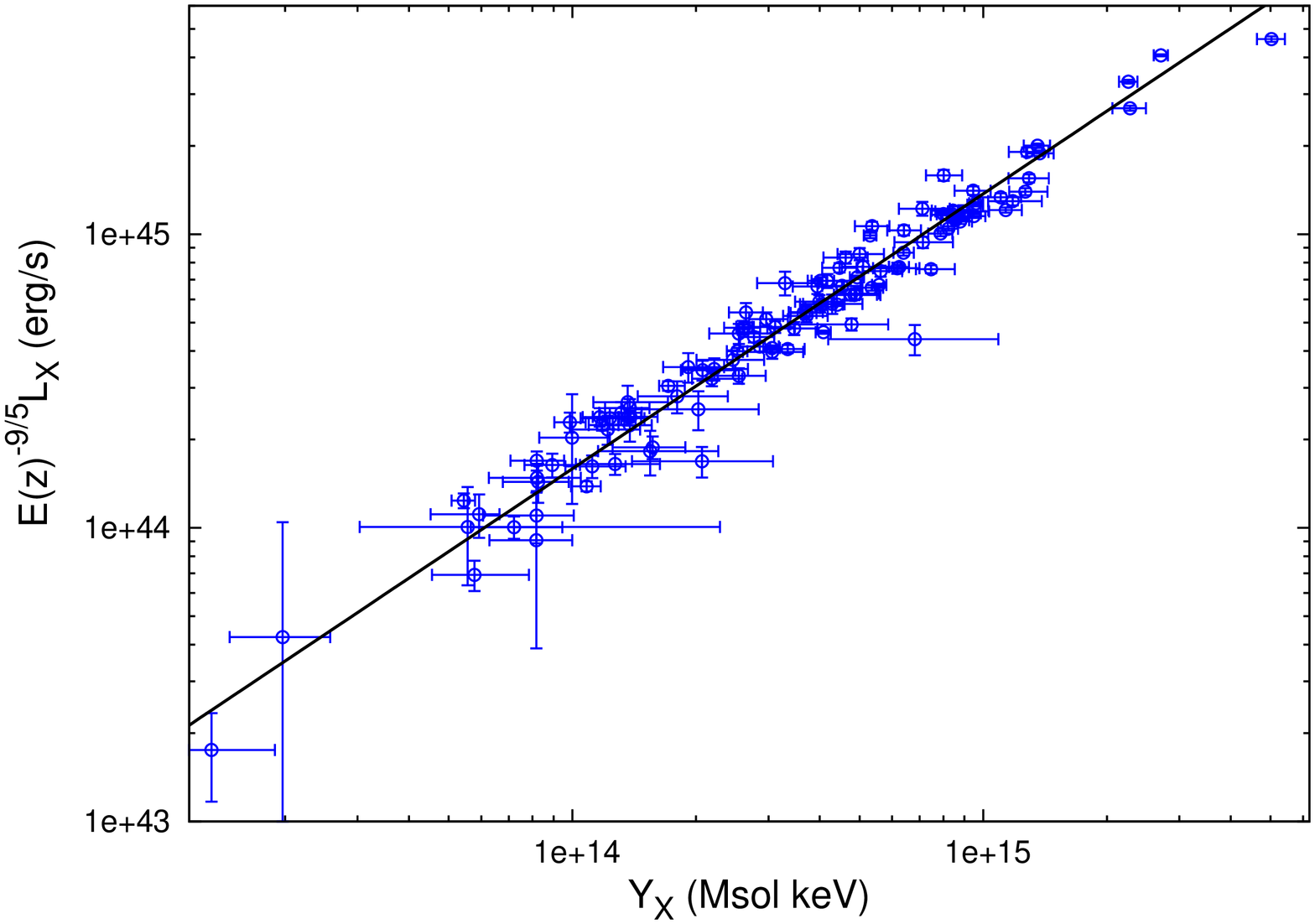,angle=270,width=6.5cm}
}}
\vspace{-4.7truecm}
\caption{Left panel: the relation between $Y_X=M_{\rm gas}T_X$ and
  $M_{500}$. Points with error bars are for Chandra observational data
  (from \protect\citealt{2007astro.ph..3661N}). Masses of
  simulated clusters have been computed by using the same estimator,
  based on the hydrostatic equilibrium, applied to the Chandra
  data. Masses of both simulated and observed clusters have been
  rescaled with redshift according to the evolution predicted by the
  self--similar model. Right panel: the relation between $Y_X$ and the
  $X$--ray luminosity, both estimated within $r_{500}$, for a set of
  clusters at $0.1<z<1.3$ extracted from the Chandra archive (from
 \protect\citealt{2007astro.ph..3504M}). Luminosities have been
  rescaled according to the redshift dependence expected from
  self--similar evolution. }
\label{fi:yx}
\end{figure}

In an attempt to provide an X--ray observable related to the pressure,
\citet{2006ApJ...650..128K} introduced the quantity
$Y_X=M_{\rm gas}T$, defined by the product of the total gas mass times the
temperature, both measured within a given aperture. With this
definition, $Y_X$ represents the X--ray counterpart of the Compton-$y$
parameter, measured from the SZ effect. By computing this quantity for
a set of simulated clusters, \citet{2006ApJ...650..128K} showed that $Y_X$ has a very tight
correlation with the cluster mass, with a remarkably small scatter of
only 8 per cent. The application to Chandra observations of a set of
nearby relaxed clusters also shows that this relation has a comparably
small scatter. Again, once masses of simulated and observed clusters
are computed by applying the same hydrostatic estimator, the
normalisation of the $Y_X$--$M$ relation from models and data closely
agree with each other (\citealt{2007astro.ph..3661N}; see
left panel of Fig.~\ref{fi:yx}).  Since mass is determined in this
case from temperature, $Y_X$ and $M$ are not independent quantities
and, therefore, the scatter in their scaling relation may be
underestimated.

\citet{2007astro.ph..3504M} computed $Y_X$ for an extended set
of clusters extracted from the Chandra archive, in the redshift range
$0.1<z<1.3$. The results of his analysis, shown in the right panel of
Fig.~\ref{fi:yx}, indicate that $Y_X$ is also tightly correlated with
the X--ray luminosity through a relation which evolves in a
self--similar way. Again, since the gas mass is obtained from the
X--ray surface brightness, $Y_X$ and $L_X$ are not independent quantities,
thus possibly leading to an underestimate of the intrinsic scatter in
their scaling relation.

\subsection{The gas mass fraction}
The measurement of the baryon mass fraction in nearby galaxy clusters
has been recognised for several years to be a powerful method to
measure the cosmological density parameter (e.g., \citealt{1993Natur.366..429W,1999ApJ...517..627M}), while its redshift evolution
provides constraints on the dark energy content of the Universe (e.g., \citealt{2002MNRAS.334L..11A,2003A&A...398..879E}, and references therein).  Since
diffuse gas dominates the baryon budget of clusters, a precise
measurement of the ICM total mass represents a fundamental step in the
application of this cosmological test. While this method relies on the
basic assumption that all clusters contain baryons in a cosmic
proportion, a number of observational evidences show that the gas mass
fraction is smaller in lower temperature systems
\citep{2003ApJ...591..749L,2003MNRAS.340..989S}. This fact forces one
to restrict the application to the most massive and relaxed systems.
In addition, since X--ray measurements of the gas mass fraction are
generally available only out to a fraction of the cluster virial
radius, the question then arises as to whether the gas fraction in
these regions is representative of the cosmic value. Indeed,
observational evidence has been found for an increase of the gas
mass fraction with radius (e.g., 
\citealt{2003A&A...403..433C}).

In this respect, hydrodynamical simulations offer a way to check how
the gas mass is distributed within individual clusters and as a
function of the cluster mass, thus possibly providing a correction
for such biases. Indeed, \citet{2004MNRAS.353..457A} resorted
to the set of SPH clusters simulated by \citet{1998ApJ...503..569E} to calibrate the correction factor
that one needs to apply to extrapolate the baryon fraction computed at
$r_{2500}$, which is the typical radius at which it is measured, to
the cosmic value. However, since this set of simulations does not
include the effects of cooling, star formation and feedback heating,
it is not clear whether they provide a reliable description of the gas
distribution within real clusters.

\begin{figure}
\centerline{
\hbox{
\psfig{figure=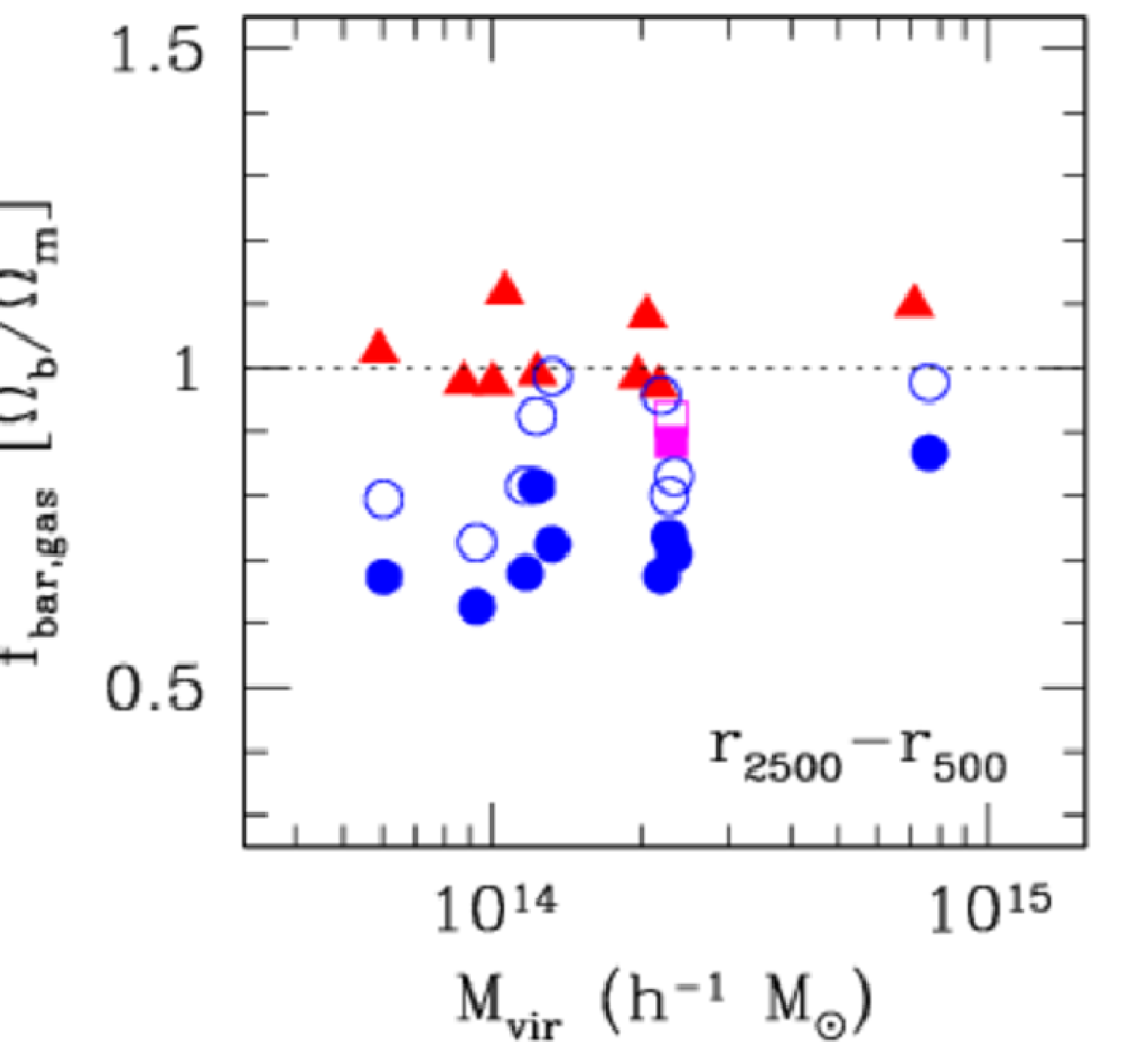,width=6.5cm}
\psfig{figure=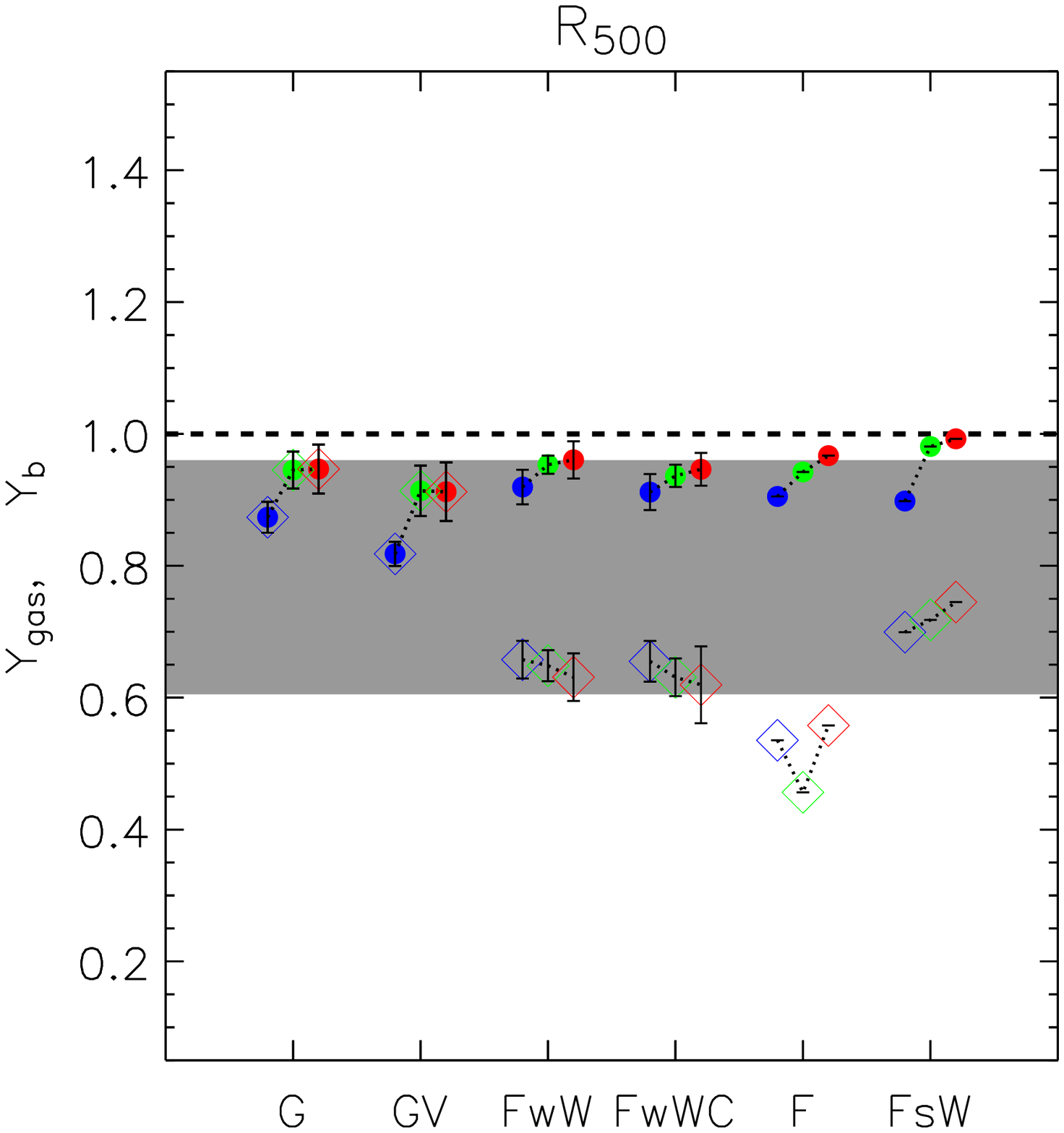,width=6.5cm}
}}
\caption{Left panel: the gas fraction (filled circles) and the baryon
  fraction (open circles) for cluster simulations which include
  cooling and star formation. Squares indicate runs where cooling is
  switched off at $z<2$, while triangles are for non--radiative
  simulations (from  \protect\citealt{2005ApJ...625..588K}). Right panel: the gas
  fraction (diamonds) and the baryon fraction (circles) within
  $r_{500}$ for SPH simulations of clusters, which include different
  gas physics. G: gravitational heating only; GV: G with a scheme for
  reduced SPH gas viscosity; FwW: radiative runs with feedback and
  weak galactic winds; FwWC: like FwW with also thermal conduction,
  with an efficiency of one--third of the Spitzer value; F: radiative
  runs with no winds; FsW: radiative runs with strong winds. Blue,
  green and red symbols refer to redshifts $z=0$, 0.7 and 1 (from  \protect\citealt{2006MNRAS.365.1021E}). The shaded area indicates
  the range of values of the observed gas fraction
  \protect\citep{1999MNRAS.305..834E}. }
\label{fi:fgas}
\end{figure}

\citet{2005ApJ...625..588K} used high resolution
simulations, based on an Eulerian code, for a set of clusters using
both non--radiative and radiative physics. They found that including
cooling and star formation has a substantial effect on the total
baryon fraction in the central cluster regions, where it is even
larger than the cosmic value. As shown in the left panel of Fig.~\ref{fi:fgas}, at the virial radius the effect is actually inverted,
with the baryon fraction of radiative runs lying below that of the
non--radiative runs. In addition, they also compared results obtained
from Eulerian and SPH codes and found small, but systematic,
differences between the resulting baryon fractions.

\citet{2006MNRAS.365.1021E} performed a similar test,
based on SPH simulations, but focusing on the effect of changing in a
number of ways the description of the relevant physical processes,
such as gas viscosity, feedback strength and thermal conduction. The
results of their analysis, which is shown in the right panel of
Fig.~\ref{fi:fgas}, demonstrated that the baryon fraction is generally
stable but only at rather large radii, $\gtrsim r_{500}$. They also
showed that changing the description of the relevant ICM physical
processes changes the extrapolation of the baryon fraction from the
central regions, relevant for X--ray measurements, while also slightly
affecting the redshift evolution.

On the one hand, these results show that simulations can be used as
calibration instruments for cosmological applications of galaxy
clusters. On the other hand, they also demonstrate that for this
calibration to reach the precision required to constrain the dark
energy content of the Universe, one needs to include in simulations
the relevant physical processes which determine the ICM observational
properties.

\section{Profiles of X--ray observables}
\label{profiles}

\subsection{The temperature profiles}
Already ASCA observations, despite their modest spatial resolution,
have established that most of the clusters show significant departures
from isothermality, with negative temperature gradients characterised
by a remarkable degree of similarity, out to the largest radii sampled
(e.g.,  \citealt{1998ApJ...503...77M}). Besides
confirming the presence of these gradients, Beppo--SAX observations
(e.g.  \citealt{2002ApJ...567..163D}) showed that
they do not extend down to the innermost cluster central regions,
where instead an isothermal regime is observed, possibly followed by a
decline of the temperature towards the centre, at least for relaxed
clusters. The much improved sensitivity of the Chandra satellite
provides now a more detailed picture of the central temperature
profiles (e.g.,  \citealt{2005ApJ...628..655V,2007arXiv0705.3865B}). At the same time, a number of
analyses of XMM--Newton observations now consistently show the
presence of a negative gradient at radii $\gtrsim 0.1r_{200}$ (e.g.,
\citealt{2005A&A...433..101P,2007A&A...461...71P}, and references therein). Relaxed
clusters are generally shown to have a smoothly declining profile
toward the centre, reaching values which are about half of the overall
virial cluster temperature in the innermost sampled regions, with
non--relaxed clusters having, instead, a larger variety of temperature
profiles.
The emerging picture suggests that gas cooling is responsible for the
decline of the temperature in the central regions, while
some mechanism of energy feedback should be responsible for preventing
overcooling, thereby suppressing the mass deposition rate and the
resulting star formation.

As for hydrodynamical simulations, they have shown to be generally
rather successful in reproducing the declining temperature profiles
outside the core regions (e.g., \citealt{2002ApJ...579..571L,2006MNRAS.373.1339R}), where gas cooling is
relatively unimportant. On the other hand, as we have already
discussed in Sect.~\ref{s:heatcool}, including gas cooling has the
effect of steepening the $T$-- profiles in the core regions, in clear
disagreement with observations. The problem of the central temperature
profiles in radiative simulations has been consistently found by
several independent analyses (e.g. 
\citealt{2003MNRAS.339.1117V,2004MNRAS.348.1078B,2007astro.ph..3661N}) and is interpreted as due to the
difficulty that currently implemented feedback schemes have in
balancing the cooling runaway.

\begin{figure}
\hbox{
\psfig{figure=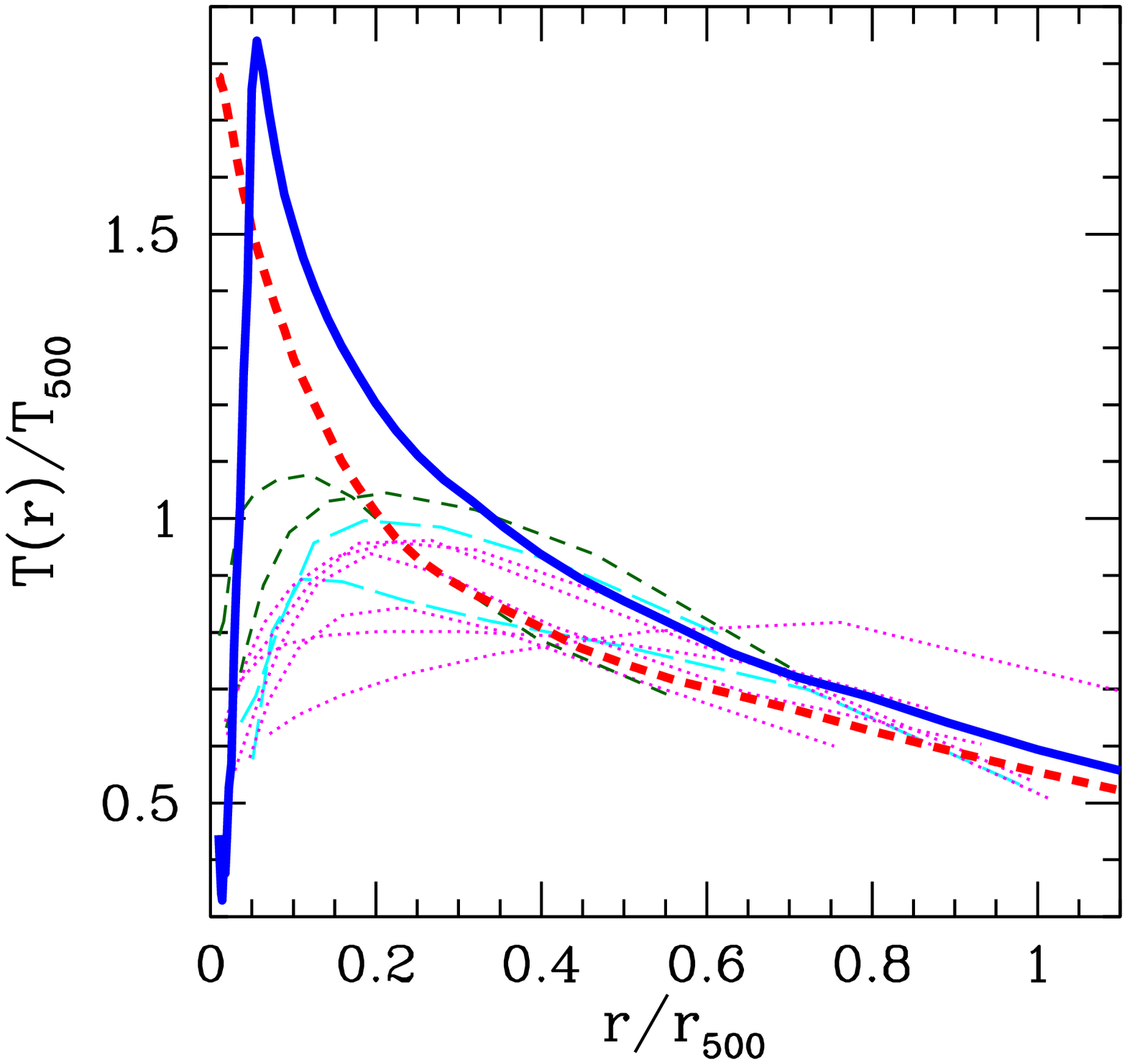,height=7cm}
\psfig{figure=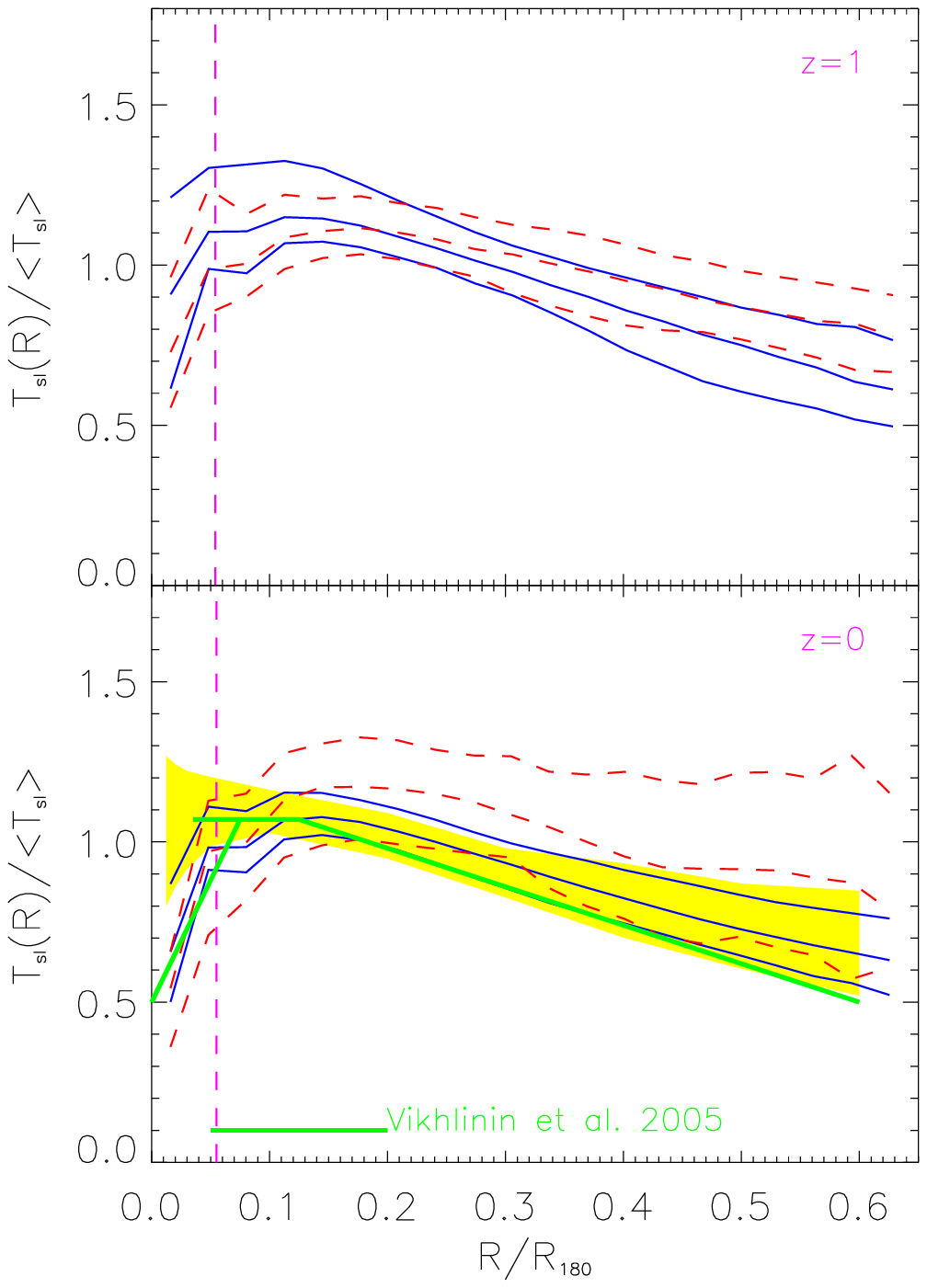,height=7cm}
}
\caption{Left panel: the temperature profiles in AMR simulations of
  clusters (thick curves) and in real clusters (thin curves). The
  simulation results are the average over 16 numerical clusters. The
  solid line is for runs with cooling and star formation, while the
  dashed line is for non--radiative simulations. The observational
  curves are for clusters with different temperatures (from
  \protect\citealt{2007astro.ph..3661N}). Right panel:
  Temperature profiles at $z=1$ (top) and at $z=0$ (bottom) from a set
  of SPH simulated clusters including cooling, star--formation and a
  ``targeted'' scheme of SN feedback (from \protect\citealt{2007MNRAS.377..317K}). Median and 10/90
  percentiles are shown. Solid and dashed lines are for irregular and
  regular clusters, respectively. The vertical dashed line indicates
  the smallest scale which is numerically resolved. In the bottom
  panel, the heavy green line shows the profile from Chandra
  observations of cool core clusters
  \protect\citep{2005ApJ...628..655V}, while the yellow shaded area is
  for a XMM--Newton sample of nearby clusters
  \protect\citep{2007A&A...461...71P}. }
\label{fi:tprofs}
\end{figure}

As an example, we show in the left panel of Fig.~\ref{fi:tprofs} the
comparison between simulated and observed temperature profiles,
recently presented by \citet{2007astro.ph..3661N},
which is based on a set of clusters simulated with an Adaptive Mesh
Refinement (AMR) code. This plot clearly shows that the central
profiles of simulated clusters are far steeper than the observed ones,
by an amount which increases when cooling and star formation are
turned on. Although this result is in qualitative agreement with other
results based on SPH codes, an eye-ball comparison with the left panel
of Fig.~\ref{fi:tsim} shows a significant difference of the
profiles in the central regions for the non--radiative runs. Indeed,
while SPH simulations generally show a flattening at $r\lesssim 0.1
r_{\rm vir}$, Eulerian simulations are instead characterised by
continuously rising profiles. While this difference is reduced when
cooling is turned on, it clearly calls for the need of performing
detailed comparisons between different simulation codes, in the spirit
of the Santa Barbara Cluster Comparison Project (e.g.  \citealt{1999ApJ...525..554F}), and to understand in detail the
reason for these differences.

As discussed above, the steep temperature profiles predicted in the
central regions witness the presence of overcooling. Viceversa, the
fact that a feedback mechanism is able to produce the correct
temperature profiles does not guarantee in itself that overcooling is
prevented.  Indeed, \citet{2007MNRAS.377..317K} found
that their ``targeted'' scheme of SN feedback produces temperature
profiles which are in reasonable agreement with observations (see the
right panel of Fig.~\ref{fi:tprofs}). However, even with this
efficient feedback scheme, the resulting stellar fraction within
clusters was still found to be too high, $\gtrsim 25$ per cent, thus
indicating the presence of a substantial overcooling in their
simulations.

Clearly, resolving the discrepancy between observed and simulated
central temperature profiles requires that simulations are able to
produce the correct thermal structure of the observed ``cool
cores''. This means that a suitable feedback should compensate the
radiative losses of the gas at the cluster centre, while keeping it at
about $\sim 1/3$ of the virial temperature. A number of analyses
converge to indicate that AGN should represent the natural solution to
this problem. Considerable efforts have been spent to investigate how
cooling can be self--consistently regulated by feedback from a central
AGN in static cluster potentials (e.g., \citealt{2001ApJ...554..261C,2004MNRAS.348.1105O,2006ApJ...643..120B,2007ApJ...656L...5S}, and reference therein), while only
quite recently these studies have been extended to clusters forming in
a cosmological context (e.g., \citealt{2006MNRAS.373L..65H,2007arXiv0705.2238S}). Although the results of these
analyses are quite promising, we still lack for a detailed comparison
between observational data and an extended set of cosmological
simulations of clusters, convincingly showing that AGN feedback is
able to provide the correct ICM thermal structure for objects spanning
a wide range of masses, from poor groups to rich massive clusters.

\subsection{The entropy profiles}
As already mentioned in Sect.~\ref{models}, a convenient way of
characterising the thermodynamical properties of the ICM is through
the entropy, which, in X--ray cluster studies, is usually defined as
$S=T/n_{\mathrm e}^{2/3}$ (see the review by  \citealt{2005RvMP...77..207V},
for a detailed discussion about the role of entropy in cluster
studies). Since the current numerical description of the
heating/cooling interplay in the central cluster regions is unable to
reproduce the observed temperature structure, there is no surprise
that simulations have difficulties also in accounting for the observed
entropy structure. However, while the ICM thermodynamics is
sensitive to complex physical processes in the core regions,
one expects simulations to fare much better in the outer regions, say
at $r>0.2\,r_{200}$, where the gas dynamics should be dominated by
gravitational processes. This expectation is also supported by the 
agreement between the observed and the simulated slope of the
temperature profiles outside cluster cores.

If gravity were the only process at work, then the prediction of the
self--similar model is that $S\propto T$. Therefore, by plotting profiles
of the reduced entropy, $S/T$, one expects them to fall on the top of
each other for clusters of different temperatures. However,
observations revealed that this is not the case.  Indications from
ASCA data \citep{2003MNRAS.343..331P} showed that entropy profiles of
poor clusters and groups have an amplitude which is higher than that
expected for rich clusters from the above scaling argument. This
result has been subsequently confirmed by the XMM--Newton data
analysed by \citet{2005A&A...429..791P} and by
\citet{2005A&A...433..101P}. Quite remarkably, a
relatively higher entropy for groups is found not only in the central
regions, where it can arise as a consequence of the heating/cooling
processes, but extends to all radii sampled by X--ray observations,
out to $\lesssim 0.5r_{500}$. These analyses consistently indicate that
$S\propto T^\alpha$, with $\alpha\simeq 0.65$, instead of $\alpha=1$
as expected from self--similar scaling.

\citet{2003MNRAS.343..331P} and \citet{2003ApJ...593..272V} interpreted this entropy excess in
poor clusters and groups as the effect of entropy amplification
generated by shocks from smoothed gas accretion. The underlying idea
is the following. In the hierarchical scenario for structure
formation, a galaxy group is expected to accrete from relatively
smaller filaments and merging sub--groups than a rich cluster
does. Suppose now to heat the gas with a fixed amount of specific
energy (or entropy). Such a diffuse heating will be more effective to
smooth the accretion pattern of a group than that of a rich cluster,
just as a consequence of the lower virial temperature of the
structures falling into the former object. In this case, accretion
shocks take place at a lower density and, therefore, are more
efficient in generating entropy. While predictions from the
semi--analytical approach by \citet{2003ApJ...593..272V}
are in reasonable agreement with observational results, one may wonder
whether these predictions are confirmed by full hydrodynamical
simulations.

\begin{figure}
\centerline{
\hbox{
\psfig{figure=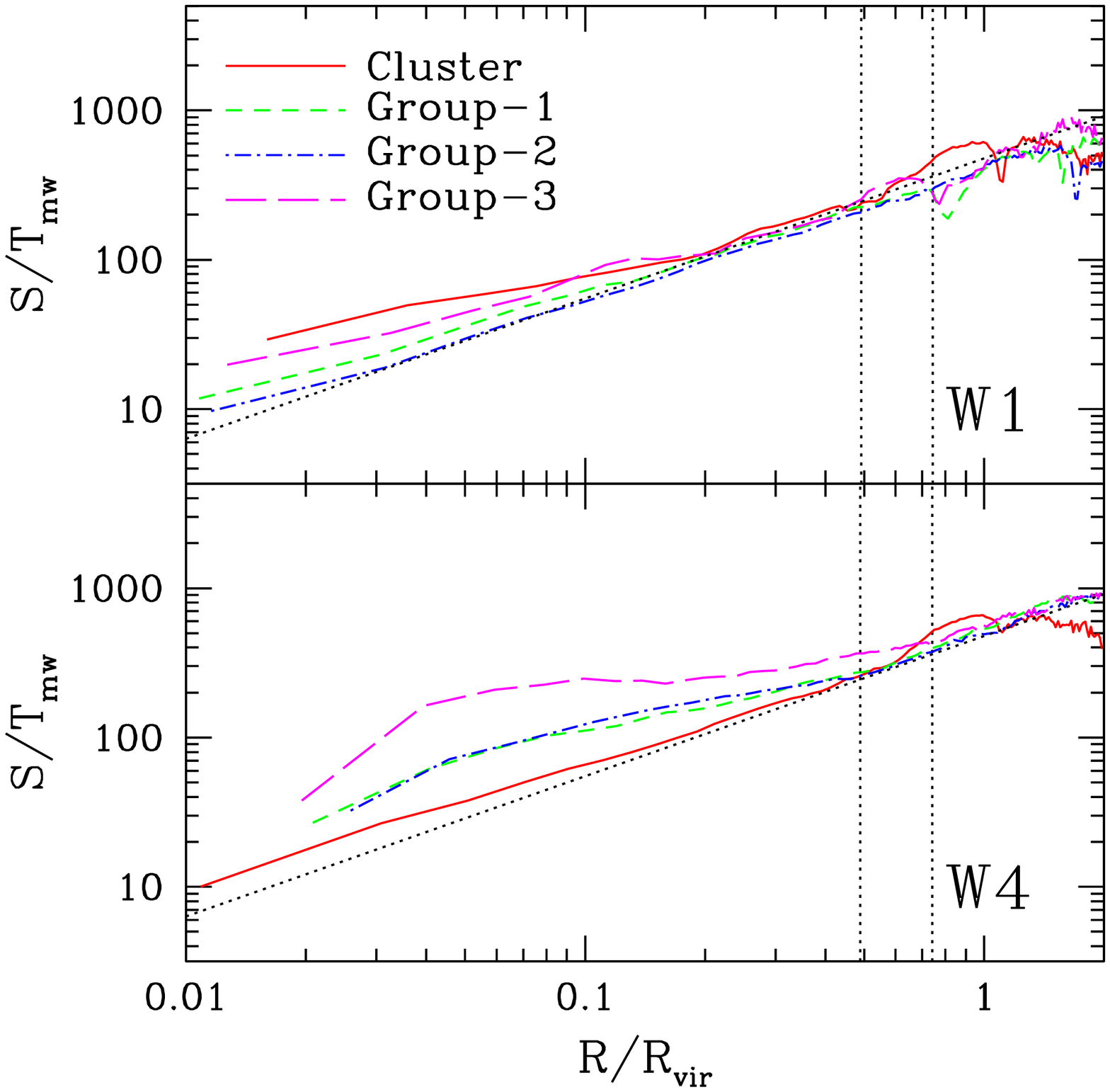,height=6.2cm}
\psfig{figure=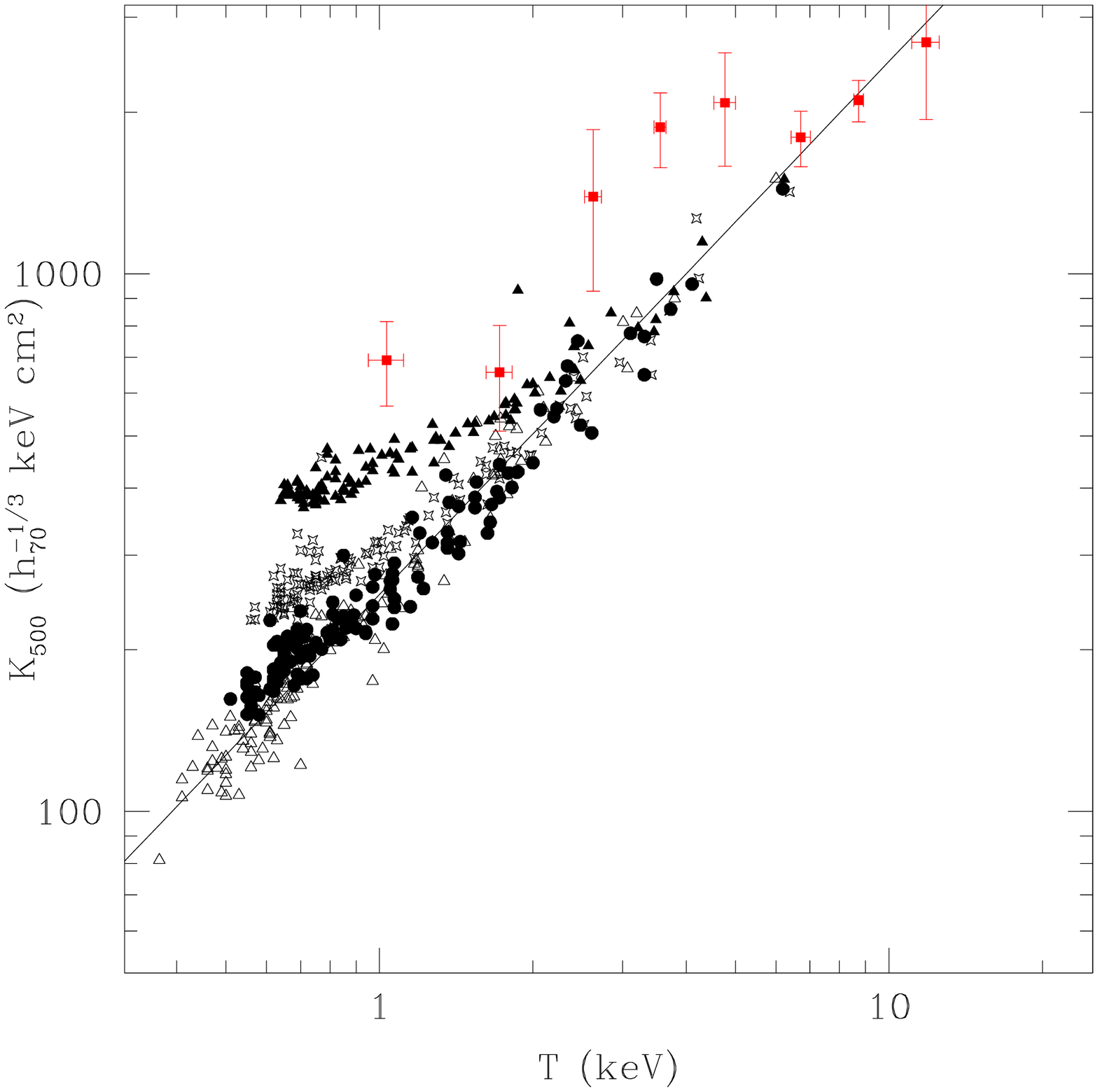,height=6.cm}
}}
\vspace{-0.2truecm}
\caption{Left panel: the profile of reduced entropy, $S/T$, for a
  series of SPH {\tt GADGET} runs of one cluster and three groups,
  including cooling, star formation and feedback from galactic
  winds. The top panel is for runs in which the winds have a velocity
  of about $360~\vel$, while the bottom panel is for an extreme wind
  model with a velocity of $830~\vel$ (from  \protect\citealt{2005MNRAS.361..233B}). The straight dotted line
  marks the slope $S\propto R^{0.95}$ which is the best--fitting to
  the observed entropy profiles
  \protect\citep{2005A&A...429..791P,2005A&A...433..101P}. Right panel:
  the relation between entropy computed at $r_{500}$ for clusters and
  groups identified in AMR ENZO cosmological simulations (from \protect\citealt{2007arXiv0705.3465Y}), compared with the
  observational results by \protect\citet{2003MNRAS.343..331P}; symbols with
  error bars. Shown are four different cases of entropy injection at
  $z=10$: no pre-heating (open triangles), 78 keV cm$^2$ (filled
  circles), 155 keV cm$^2$ (stars), 311 keV cm$^2$ (filled
  triangles). The solid line is a power--law fit to the self--similar
  prediction from the simulations.}
\label{fi:entrpr}
\end{figure}

To tackle this problem, \citet{2005MNRAS.361..233B}
performed a series of SPH hydrodynamical simulations of four galaxy
clusters and groups, with temperature in the range 0.5--3 keV, using
both non--radiative and radiative runs, also exploring a variety of
heating recipes. As a result, they found that galactic ejecta powered
by an even extremely efficient SN feedback are not able to generate
the observed level of entropy amplification. This result is shown in
the left panel of Fig.~\ref{fi:entrpr} where the profiles of reduced
entropy are plotted for the simulated structures in the case of
moderate (upper panel) and very strong (bottom panel) galactic
outflows. Borgani et al. also showed that only adding an entropy floor
at $z=3$ provides an efficient smoothing of the gas accretion pattern
and, therefore, a substantial degree of entropy amplifications (see
Fig.~\ref{fi:csf}). Although this diffuse pre--heating provides an
adequate entropy amplification at large radii, the price to pay is
that the entropy level is substantially increased in the central
regions. This is at variance with high--resolution Chandra
measurements of low entropy gas in the innermost cluster regions,
where it reaches values as low as $\sim 10$ keV cm$^2$ \citep{2006ApJ...643..730D}.

A similar analysis has also been performed by \citet{2007arXiv0705.3465Y}, who used a set of AMR non--radiative
cosmological simulations performed with the ENZO code
\citep{2004astro.ph..3044O}, with gas entropy boosted at high redshift,
$z=10$. In keeping with previous analyses, they found that pre--heated
simulations are generally able to reproduce the observed
luminosity--temperature and mass--temperature relations. However,
differently from \citet{2005MNRAS.361..233B}, even
the most extreme pre--heating scheme does not provide an appreciable
degree of entropy amplification. This result is shown in the right
panel of Fig.~\ref{fi:entrpr}. The deviations from self--similarity
for the entropy computed at $r_{500}$ are always too small to
reconcile the simulations with the observational data by \citet{2003MNRAS.343..331P}. A possible reason for the different
results with respect to the analysis by Borgani et al. is due to the
higher redshift of pre--heating, $z=10$ instead of 3, used by Younger
\& Bryan. Furthermore, grid--based codes are known to produce entropy
profiles that, in the central part of the halos, $r\lesssim 0.2
r_{200}$, are flatter than those produced by SPH codes (e.g.,  \citealt{2005MNRAS.364..909V}). Therefore, the question arises as to
whether different numerical schemes of hydrodynamics react in
different ways to pre--heating. It would be highly recommendable that
future simulation comparison projects will include tests of how
different codes behave in the presence of simple schemes of
non-gravitational heating.

\section{Summary}
\label{summary}
In this paper we reviewed the current status of the comparison
between cosmological hydrodynamical simulations of the X--ray
properties of the intra--cluster medium (ICM) and observations. We
first presented the basic predictions of the self--similar model,
based on the assumption that only gravitational processes drive the
evolution of the ICM. We then showed how a number of observational
facts are at variance with these predictions, with poor clusters and
groups characterised by a relatively lower density and higher entropy
of the gas. This calls for the need of introducing some extra physical
processes, such as non--gravitational heating and radiative cooling,
which are able to break the self--similarity between objects of
different size. The results based on simulations, which include these
effects, can be summarised as follows.\\ {\bf (1)} The observed
scaling relation between X--ray luminosity and temperature can be
reproduced by simulations including cooling only, but at the expense
of producing a too large fraction of cooled gas. Introducing ad--hoc
schemes of entropy injection at high redshift (the so--called
pre--heating) can eventually produce the correct $L_X$--$T$
relation. However, no simulations have been presented so far, in which
a good agreement with observations is achieved by using a feedback
scheme in which the energy release and thermalisation from SN or AGN
is self--consistently computed by the simulated star formation or
accretion onto a cosmologically evolving population of black holes.\\
{\bf (2)} Simulations which include cooling, star formation and SN
feedback, correctly reproduce the observed mass--temperature and
``pressure''--temperature relations. Quite interestingly, this
agreement is achieved once the masses of simulated clusters are
estimated by applying the same mass estimators, based on the
assumption of hydrostatic equilibrium, which are applied to the
analysis of real clusters. The violation of hydrostatic equilibrium,
related to the non--thermal pressure support from subsonic gas
motions, would otherwise lead to a $\sim 20$ per cent overestimate of
the normalisation of the above scaling relations for simulated
clusters.\\ {\bf (3)} Both non--radiative and radiative runs naturally
predict the observed negative gradients of the temperature profiles
outside the cluster core regions, $r\gtrsim 0.2 r_{200}$. This
suggests that the ICM thermal structure in these regions is indeed
dominated by the action of gravitational processes, such as heating
from shocks associated to supersonic gas accretion.\\ {\bf (4)}
Introducing cooling has the effect of steepening the temperature
profiles in the innermost regions, as a consequence of the adiabatic
compression of inflowing gas, caused by the lack of pressure
support. Therefore, gas in the cores of simulated clusters generally
lies in a high--temperature phase, which is very well separated from
the cold ($\sim 10^4$ K) phase. This is at variance with the observed
``cool core'' structure of real clusters, which show instead the
presence of a fair amount of gas, down to a limiting temperature of
$\sim 1/3$ of the virial temperature, which formally has a rather
short cooling time. The presence of this gas makes the observed
temperature profiles of relaxed clusters to peak at $r\simeq
0.2r_{200}$, thereby decreasing by about a factor two in the innermost
sampled regions. The discrepancy between the simulated and the
observed ``cool core'' calls for the need of introducing in cluster
simulations a suitable feedback mechanism which is able to compensate
the radiative losses, keeps pressurised gas in the central regions,
and suppresses the mass deposition rate.\\ {\bf (5)} Simulations are
generally rather successful in reproducing the observed slope of the
entropy profiles outside the core regions, $S\propto r^\alpha$ with
$\alpha \simeq 1$. This slope is also close to that predicted by
models based on gravitational heating from spherical accretion, thus
lending further support to the picture that gravity drives the
evolution of the ICM outside the core regions. Quite intriguingly,
however, the normalisation of the profiles out to the largest sampled
radii, $\lesssim r_{500}$, is observed to have a milder temperature
dependence than expected from the self--similar model, $S\propto
T^\alpha$ with $\alpha\simeq 0.65$ instead of $\alpha\simeq 1$. This
entropy excess in poor systems can be reproduced in simulations only
by adding a diffuse pre--heating mechanism, which smoothes the pattern
of gas accretion, thereby leading to an amplification of entropy
generation from shocks in poorer systems. However, this mechanism is
able to provide the correct explanation only for clusters with
temperature $T\lesssim 3$ keV.  Generating entropy amplification in
hotter systems would require an exceedingly large amount of
pre--heating, which would generate too shallow entropy profiles.

In general, the above results demonstrate the capability of
cosmological hydrodynamical simulations to predict the correct
thermodynamical properties of the ICM outside the central regions of
galaxy clusters. On the other hand, even simulations based on an
efficient SN feedback fail in preventing overcooling and providing
the correct description of the thermodynamical ICM properties in the
central regions. The ``cool core'' failure of simulations based on
stellar feedback is also witnessed by the exceedingly massive and blue
central galaxies that are generally produced (see also \citealt{borgani2008} - Chapter 18, this volume).

The fact that cool cores are observed for systems spanning at least
two orders of magnitude in mass, from $\sim 10^{13}M_\odot$ to $\sim
10^{15}M_\odot$, suggests that only one self--regulated mechanism
should be mainly responsible for this, rather than the combination of
several mechanisms possibly acting over different
time--scales. Although AGN are generally thought to naturally provide
such a mechanism, a detailed comparison between observational data and
an extended set of cluster simulations, including this kind of
feedback is still lacking. Furthermore, understanding in detail the
role of AGN in determining the ICM properties requires addressing in
detail two major issues. Firstly, the accretion process onto the
central black hole involves scales of the order of the parsec, while
the X--ray observable effects involve scales of $\sim 10$--100
kpc. This requires understanding the cross--talk between two ranges of
scales, which differ by at least four orders of magnitude. Secondly,
the total energy budget available from AGN is orders of magnitude
larger than that required to regulate cooling flows. Therefore, one
needs ultimately to understand the channels for the thermalisation of
the released energy and how this naturally takes place without
resorting to any ad--hoc tuning. These problems definitely need to be
addressed by simulations of the next generation, which will be aimed
at understanding in detail the evolution of the cosmic baryons and the
observational X--ray properties of galaxy clusters.

\begin{acknowledgements}
 The authors thank ISSI (Bern) for support of the
team ``Non--virialized X-ray components in clusters of galaxies''. SB
wishes to thank Hans B\"ohringer, Stefano Ettori, Gus Evrard, Alexey
Finoguenov, Andrey Kravtsov, Pasquale Mazzotta, Silvano Molendi,
Giuseppe Murante, Rocco Piffaretti, Trevor Ponman, Elena Rasia, Luca
Tornatore, Paolo Tozzi, Alexey Vikhlinin and Mark Voit for a number of
enlightening discussions. Partial support from
the PRIN2006 grant ``Costituenti fondamentali dell'Universo'' of
the Italian Ministry of University and Scientific Research
and from the INFN grant PD51 is also gratefully acknowledged.
\end{acknowledgements}

\bibliographystyle{aa}
\bibliography{13_borgani}

\end{document}